\documentclass[10pt, conference, letterpaper]{IEEEtran}
\IEEEoverridecommandlockouts
\usepackage[cmex10]{amsmath}
\usepackage{cite}
\usepackage{amsmath, amsthm, amssymb}
\usepackage{algorithm}
\usepackage[noend]{algpseudocode}
\usepackage{tabularx}
\usepackage{fancybox}
\usepackage{float}
\usepackage{subfig}
\usepackage{graphicx}
\usepackage{mathtools}
\usepackage{color}
\usepackage{array}
\newcolumntype{C}[1]{>{\centering\arraybackslash}m{#1}}

\newtheorem{theorem}{\textbf{Theorem}}

\theoremstyle{definition}

\newtheorem*{problem*}{\textbf{Problem}}
\newtheorem{definition}{\textbf{Definition}}

\newcommand{\name}{WatchDog}

\theoremstyle{definition}
\newtheorem{example}{\textbf{Example}}

\def\BibTeX{{\rm B\kern-.05em{\sc i\kern-.025em b}\kern-.08em
    T\kern-.1667em\lower.7ex\hbox{E}\kern-.125emX}}
\begin{document}

\title{WatchDog: Real-time Vehicle Tracking on \\ Geo-distributed Edge Nodes \vspace{-10 pt}}

\author{\IEEEauthorblockN{Zheng Dong\IEEEauthorrefmark{2},
Yan Lu\IEEEauthorrefmark{4},
Guangmo Tong\IEEEauthorrefmark{5},
Yuanchao Shu\IEEEauthorrefmark{3},
Shuai Wang\IEEEauthorrefmark{1},
and 
Weisong Shi\IEEEauthorrefmark{2}}
\IEEEauthorblockA{\IEEEauthorrefmark{2}Wayne State University, \IEEEauthorrefmark{3}Microsoft Research, \IEEEauthorrefmark{1}Southeast University, \IEEEauthorrefmark{4}New York University, \IEEEauthorrefmark{5}University of Delaware}\vspace{-10pt}}
\maketitle

\begin{abstract}
% Vehicle tracking as an essential smart city technology enables surveillance tasks like hit-and-run monitoring and criminal behavior localization. The ubiquitous edge devices are promising platforms for executing machine learning algorithms used in tracking tasks, but their limited local computing resources can result in a significant processing delay that leads to tracking failures. In coping with such challenges, we this paper proposes a real-time tracking system, WatchDog, a co-design of machine learning methods and real-time task scheduling. We design a hierarchical machine learning framework for vehicle detection and re-identification with an adjustable inference time,  which enables a flexible real-time schedule such that the real-time correctness can be guaranteed. In particular, WatchDog can automate the selection of machine learning methods, and it ensures a provable response time bound at each edge device. Extensive evaluations on WatchDog have been conducted using our accessible real-world vehicle system-wide datasets. Experimental results show that WatchDog can guarantee 100\% tracking coverage of the VoI in real-time.

Vehicle tracking, a core application to smart city video analytics, is becoming more widely deployed than ever before thanks to the increasing number of traffic cameras and recent advances of computer vision and machine learning. Due to the constraints of bandwidth, latency, and privacy concerns, tracking tasks are more preferable to run on edge devices sitting close to the cameras. However, edge devices are provisioned with a fixed amount of compute budget, making them incompetent to adapt to time-varying and imbalanced tracking workloads caused by traffic dynamics. In coping with this challenge, we propose WatchDog, a real-time vehicle tracking system fully utilizes edge nodes across the road network. WatchDog leverages computer vision tasks with different resource-accuracy trade-offs, and decomposes and schedules tracking tasks judiciously across edge devices based on the current workload to maximize the number of tasks while ensuring a provable response time bound at each edge device. Extensive evaluations have been conducted using real-world city-wide vehicle trajectory datasets, showing a $100\%$ tracking coverage with real-time guarantee.
\end{abstract}

%!TEX root = main.tex

\section{Introduction}
Smart city traffic safety initiatives are springing up across the world as more cities embrace big data and video analytics. A straightforward solution of smart city data analytics is to aggregate data and conduct centralized processing in the cloud. This paradigm, however, has several downsides. First, video data uploading requires a substantial amount of network bandwidth, especially under the increasing number of high resolution cameras. Second, cloud-based processing adds up latency, which could be prohibitively high for smart city applications such as amber alerts or traffic light control. Third, privacy becomes more of an issue when public information is uploaded and stored in the cloud. 
% The substantial transmission and processing delay impedes the development of real-time applications (e.g., the hit-and-run tracking system) in the surveillance systems, resulting in a utilization loss of the real-time information captured by the cameras. In coping with such challenges, this paper aims to design a real-time vehicle tracking system through decentralized data processing.

%Nowadays, increasing number of smart edge devices, such as Azure Data Box Edge~\cite{Azure}, Argonne Waggle~\cite{Waggle} and Intel Fog Reference~\cite{Fog}, are being developed and deployed in cities around the world. Such smart edge devices enable fast and local computation, and  have emerged to integrate smart city services and city resources, and thus improve city performance in the domain of public safety. Security surveillance systems (e.g., the induction coil system~\cite{leduc2008road}, the real-time traffic surveillance system~\cite{bramberger2004real} and the suspicious object monitoring system~\cite{ishii2005monitor}) are deployed in busy areas to collect large amounts of vehicle monitoring information. By analyzing the newly available data, criminal behaviors, such as hit-and-run accidents, can be detected efficiently through offline programs. 

In coping with such challenges, an increasing number of smart edge devices, such as Azure Stack Edge~\cite{Azure}, Argonne Waggle and Intel Fog Reference, are being developed and deployed in cities around the world. Such smart edge devices enable fast local computation and have benefited security surveillance systems such as induction coil system~\cite{leduc2008road}, traffic surveillance system~\cite{bramberger2004real} and the suspicious object monitoring system~\cite{ishii2005monitor}. The shift in computing paradigm from the cloud to the edge has also necessitated the adoption of new programming models, algorithms, and analytics methods to fully exploit the computing capacity of multi-core chips deployed on edge. Edge node manufacturers and application developers are starting to discover ways to multiplex tasks and share resources across nodes in an edge cluster~\cite{hung2018videoedge,wang19hotedgevideo}. This advanced edge computing approaches provide new possibilities for designing real-time video analytics systems, which leverage machine learning for tasks like object detection and re-identification. In this paper, we focus on multi-camera vehicle tracking, a core application of smart city video analytics system, and study how to leverage geo-distributed edge nodes to build a reliable real-time tracking system. %The advantages of real-time tracking system are evident: (\textit{i}) vehicle information is captured and processed in real-time, and the corresponding reactions (e.g., real-time tracking) can be performed instantly; (\textit{ii}) the privacy issue does not exist since the original data is discarded at edge nodes after being processed. In this paper, we study how to leverage such promising platforms to build a reliable real-time tracking system.

%The computing industry recently experienced a major shift in computing paradigm from the cloud to the edge with the advent of multi-core chips. This shift has necessitated the adoption of new programming models, algorithms,and analysis methods to fully exploit the computing capacity of multi-core chips deployed on the edge. Edge node manufacturers and application developers are starting to discover the benefits of doing more compute and analytics on the edge nodes themselves using powerful embedded computing platforms. This advanced on-edge approach provides new possibilities of designing real-time surveillance systems, which rely on machine learning for tasks such as object recognition. The advantages of the advanced real-time surveillance system are evident: (\textit{i}) vehicle monitoring information is captured and processed in real-time, and the corresponding reactions (e.g., real-time tracking) can be performed instantly; (\textit{ii}) the privacy issue is resolved because the surveillance data is discarded at edge nodes after being processed.

\begin{figure}[t]
\centering
\subfloat[Morning rush hour]{%
\includegraphics[width=10pc]{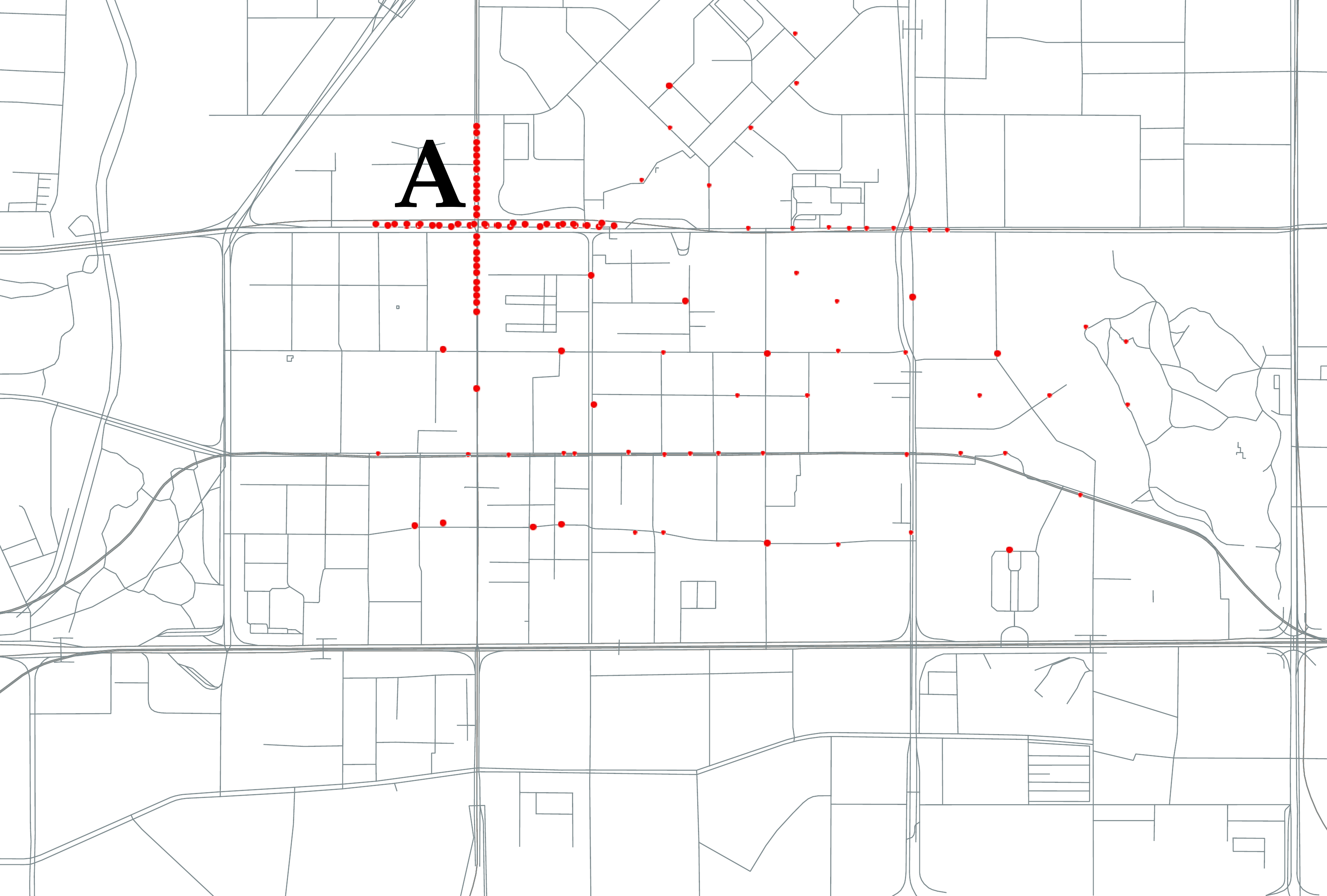}}
\label{fig:a}\hfill
\subfloat[Evening rush hour]{%
\includegraphics[width=10pc]{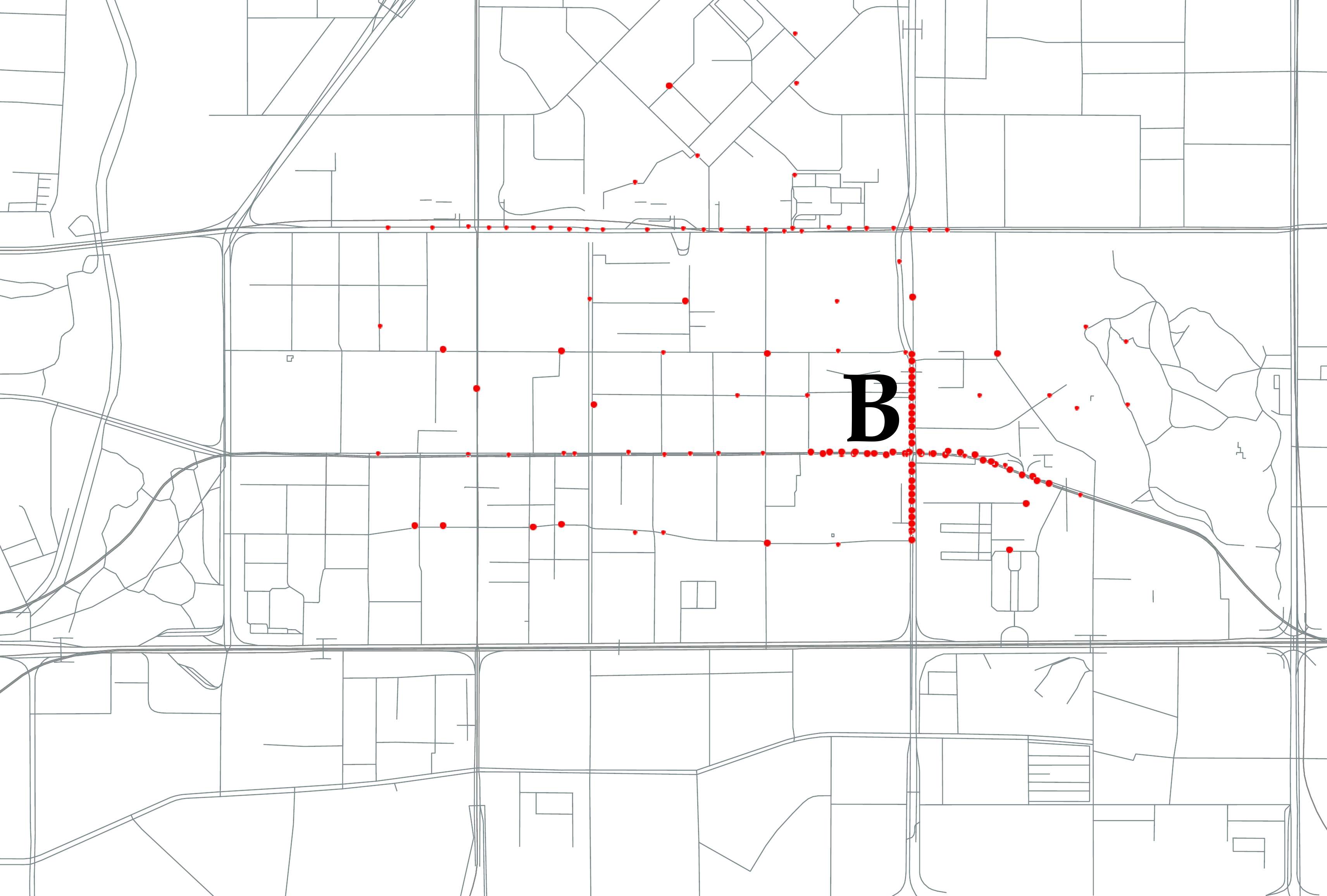}}
\label{fig:b}\hfill
\caption{\small Snapshots of traffics during different rush hours in the city of Shenzhen, China. Red dots denote vehicles traveling in this area.}
\label{fig:intro}
\vspace{-15pt}
\end{figure}

\noindent{\bf Motivation:} Edge nodes are provisioned with fixed compute budget. For instance, Azure Stack Edge~\cite{Azure} node has $2 \times 10$ core CPUs and 128 GB memory. Even though they are beefy enough to handle computational demands for real-time data processing in most of the cases, in some corner cases, when the number of vehicles appearing in the monitored areas is too large, the corresponding data processing may not be able to complete in time, leading to a fatal failure for the whole system. Fig.~\ref{fig:intro} shows the snapshots of traffic conditions during different rush hours in Shenzhen. Suppose that a real-time tacking system is implemented on the edge node at each intersection to track hit-and-run vehicles. As seen in Fig.~\ref{fig:intro}(a), at most of the intersections, only one or two vehicles appear in the monitored areas and hence the tracking system can run complex vision algorithms on each of the vehicle and identify Vehicle-of-Interest (VoI) easily. However, during morning rush hours, number of vehicles at intersection A grows substantially. Thus, identifying VoI from all vehicles traveling through intersection A in real-time requires far more computing resources, which may exceed the computing capacity of the edge node. If the VoI cannot be identified at intersection A, the tracking system will lose the VoI. Similar patterns are observed at intersection B, where the corresponding edge node falls short during evening rush hours. One the other hand, compute budget provision on all edge nodes based on the worst case scenario is also a non-starter due to the high upfront investment and low resource utilization. In light of the tension between traffic variations and fixed amount of edge compute resources, this paper aims to answer a simple question: can we collaboratively utilize the existing geo-distributed edge nodes deployed in the city to provide a reliable tracking system without ``tracking loss'' at crowded intersections? 

% A straightforward idea to enhance the performance of the tracking system is to add extra computing resources on the edge node at intersection A. However, this may not work. As seen in Fig.~\ref{fig:intro}(b), during the evening rush hour, intersection A is no longer at the main traffic thoroughfares and vehicles are crowded at intersection B. In other words, even we deploy extra computing resources on the edge node at intersection A, the VoI may be lost at intersection B. With the urban development and enlargement, the main traffic thoroughfares change over time. It is impossible to upgrade the computing platforms at all possible crowded intersections due to the huge amount of unacceptable cost and the relatively low utilization. From a different perspective, the computing capacity of the whole tracking system is not fully utilized. When the VoI gets lost at intersection A in Fig.~\ref{fig:intro}(a), all the other edge nodes in this area are idle. 

\begin{figure}[t]
\centering
\subfloat[Road Network]{%
\includegraphics[width=10pc]{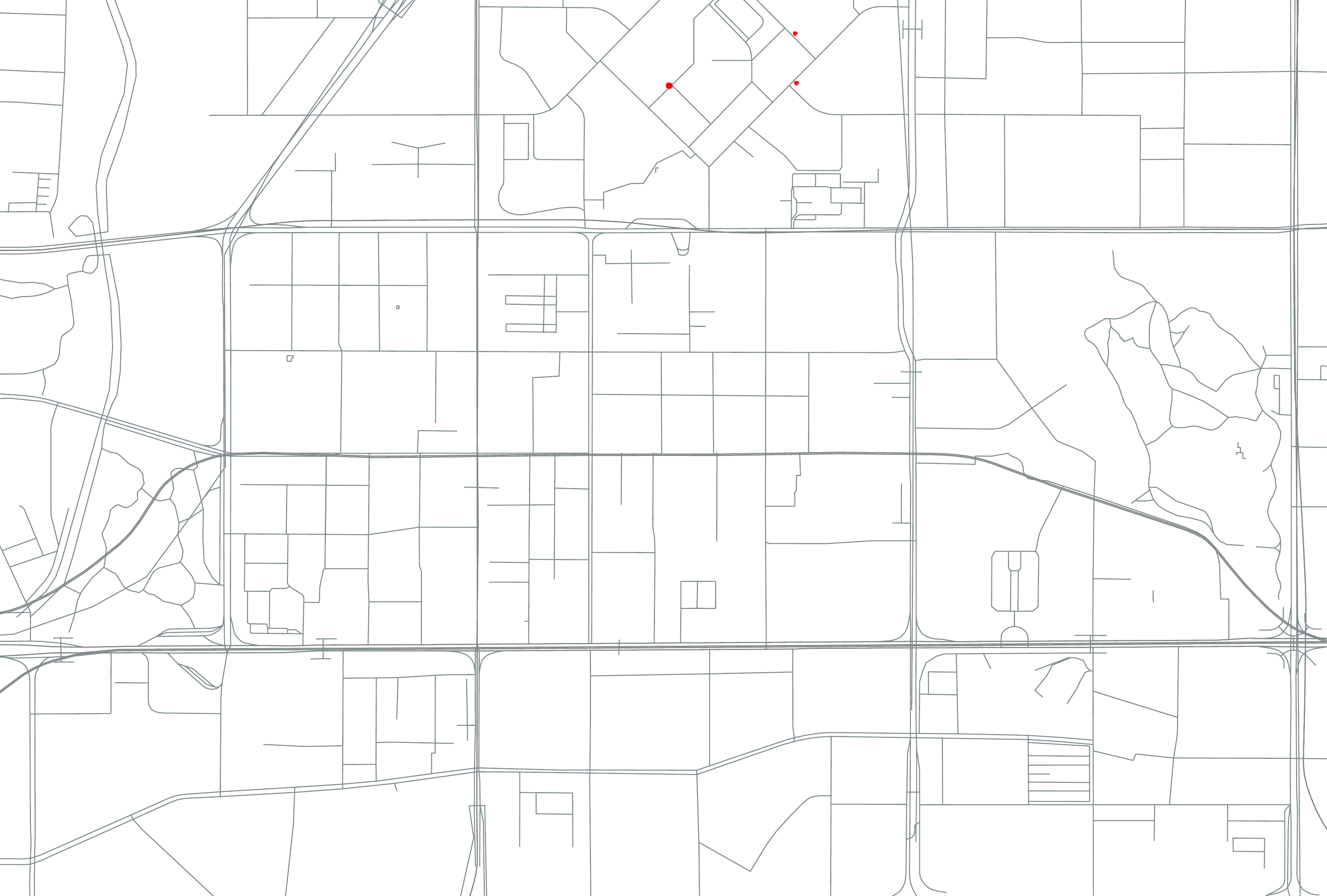}}
\label{fig:a}\hfill
\subfloat[Real-time traffic information.]{%
\includegraphics[width=10pc]{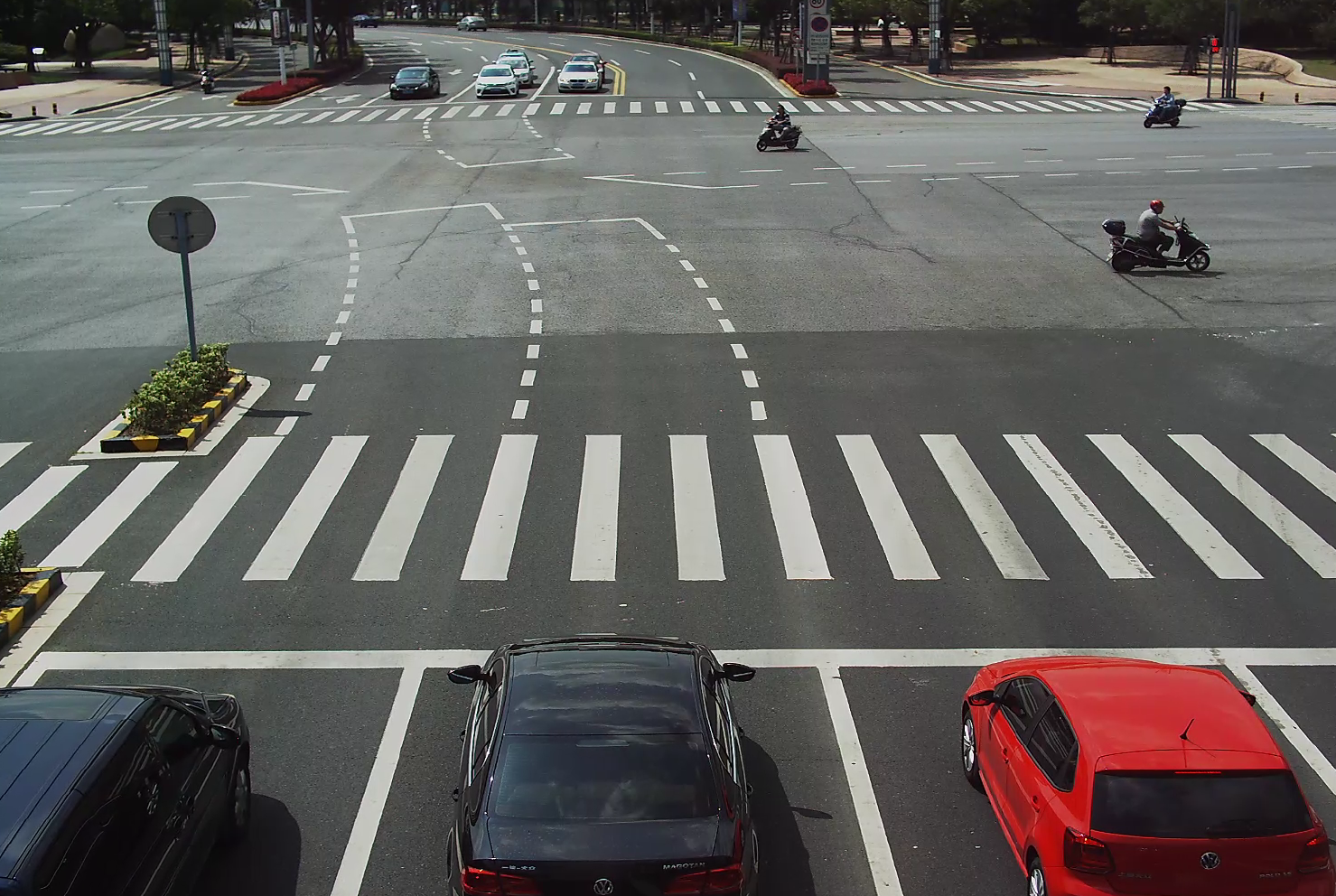}}
\label{fig:b}\hfill
\caption{\small A road network and surveillance video at an intersection.}
\label{fig:roadnetwork}
\vspace{-15pt}
\end{figure}

Inspired by the idea of collaborative tracking (i.e., the tacking task will be decomposed and scheduled judiciously across the distributed edge nodes based on real-time traffic conditions), in this paper, we propose the design of an intelligent real-time tracking system in smart cities, named \name, to track vehicles across intersections. There are two major components in \name: \emph{a real-time admission control policy} and \emph{a novel dynamic Vehicle - ReID framework}. \name\ models video analytics executed on an edge node to ReID the VoI as a real-time system: each real-time task corresponds to a vehicle re-identification module performed on a detected vehicle. The execution time of a real-time task depends on the vehicle re-identification module chosen from the dynamic Vehicle - ReID framework, and whether or not each real-time task can complete in real-time is verified by the admission control policy. The longer the tasks execute, the less chance the real-time task system has to pass the admission control. By combining the real-time admission control policy with the dynamic Vehicle - ReID framework together, \name\ builds a reliable tracking system without ``tracking loss'' at crowded intersections. 

\noindent{\bf Example:} Suppose a Mercedes silver GLB SUV (which is the VoI) is involved in a hit-and-run accident. When it enters a crowded intersection, the computation capacity on the edge node is not enough to perform the most fine-grained re-identification method to check each vehicle and identify VoI. Then the Vehicle - ReID module is downgraded to a coarse but more lightweight method to identify the vehicles' colors and models, which is determined by the admission control policy. The tracking system detects all silver SUVs at the crowded intersection, which are traveling to different neighboring intersections. The tracking system informs the edge nodes to ReID the VoI at corresponding intersections. When the silver SUVs enter the intersections where the traffic is light, the VoI will be identified through advanced matching methods and the other silver SUVs will be eliminated. Following this method, the ``tracking loss'' issue is solved. Our specific contributions are listed as follows: 

\begin{itemize}
    \item We propose a simple yet effective real-time system for tracking hit-and-run vehicles in smart cities, which for the first time enables us to collaboratively utilize the distributed edge resources in the road network to enhance the performance of the whole tracking system.
    \item To the best of our knowledge, this is the first in-depth work to investigate the combined effect of video processing latency and real-time traffic conditions, which are not discussed in the existing solutions. A real-time admission control policy and a novel dynamic Vehicle - ReID framework are proposed to resolve the ``tracking loss'' issue.
    \item We have extensively evaluated WatchDog using our accessible real-world vehicle system-wide datasets. Experimental results show that WatchDog can guarantee 100\% tracking coverage of the VoI in real-time without ``tracking loss''.
\end{itemize}

%!TEX root = main.tex

\section{Basic Setup and System Model}
\subsection{Basic Setup}
Given the road network in the urban area of a smart city, this paper aims to find a method to track the VoI (e.g., hit-and-run vehicles) by combining the information from fixed video surveillance cameras and the road network, thereby helping the tracking system track the VoI in real-time using minimum edge resources. The basic settings of this paper are outlined as follows:

\begin{itemize}
    \item {\bf Road Network:} Fig.~\ref{fig:roadnetwork} shows a road network in the urban area of Shenzhen with intersections and road segments. Surveillance cameras are deployed at the intersections to capture the real-time traffic information.
    \item {\bf Edge nodes:} Consistent with existing deployments, our focus is on ``edge'' computation of video analytics. In our setup, one edge node is deployed at each intersection, which consists of a surveillance camera and a computing platform. A video captured by the camera is streamed to this edge box and the pipeline modules including object detection and re-identification algorithms are run on this edge node.
    \item {\bf Cloud:} Each edge node only captures and processes local information at the intersections. In order to get a global view of the entire monitored area, the cloud collects the processing results from all the edge nodes. Thus, the tracking system includes both edge nodes and cloud.
\end{itemize}

\subsection{System Model}
In this subsection, we define the road network and vehicle trajectory used in our system model.

\begin{definition}
{\bf(Road Network)} The road network consists of intersections and road segments between intersections, which can be modeled as a graph $G(V, E)$ where the vertex set $V$ denotes all intersections and the edge set $E$ corresponds to all road segments. Intersection $I_i \in V$ and $e_{i,j} \in E$ if there exists a road segment from intersection $I_i$ to intersection $I_j$.
\end{definition}

\begin{definition}
{\bf(Vehicle Trajectory)} For an arbitrary vehicle, which travels in the road network, its trajectory in one specific day $d$ can be formulated by $T = \{I_o (t_1, t_2),$ $I_{o+1} (t_3, t_4),\dots\dots, I_e (t_x, t_{x+1})\}$, which means this vehicle is firstly detected at intersection $I_o$, then ends at intersection $I_e$. For one element $I_y (t_k, t_{k+1})$ in this trajectory, $t_k$ is the time instant that the vehicle enters intersection $I_y$ and time instant $t_{k+1}$ corresponds to the time it leaves this intersection.
\end{definition}

%\begin{definition}
%{\bf(Active Period)} Each intersection is covered by an edge node, which is used for processing the frames captured from the intersection. When a hit-and-run accident is reported, involved edge nodes will be activated during their \textit{active periods} within which the VoI Re-identification algorithms will be performed on each frame to identify and track the VoI at the intersections. The active periods for involved edge nodes are calculated based on the VoI's trajectory and other vehicles' historical information at this intersection. The method to calculate the active period is given in Sec.~\ref{sec:watchdog}. After the active periods, the captured videos will be discarded.
%\end{definition}

\begin{definition}
{\bf(The Tracking System)} The tracking system includes both the edge nodes and the cloud. The object recognition algorithms are implemented on the edge nodes. At the very beginning, when a hit-and-run accident is reported to the tracking system, the first edge node is activated instantly according to the location of the accident. During the \emph{active period} of the edge node, it identifies the VoI and the next intersection on the VoI's trajectory based on real-time video analytics. The analysis results are uploaded to the cloud to active the edge node deployed at the next intersection and stop the previous one. From then on, the next activated edge node and its active period depend on the analysis results performed at the previous intersection and the historical traffic information (which will be discussed in section~\ref{sec:watchdog}). The communication between edge nodes is enabled through the cloud.    
\end{definition}

Intuitively, if the tracking system knows when the VoI will arrive at which intersections in advance, the corresponding edge nodes can be activated during specific periods to track the VoI in real-time. However, in practice, the implementation of real-time tracking is very challenging. 

\begin{figure}[t!]
\centering
\includegraphics[width=2.5in]{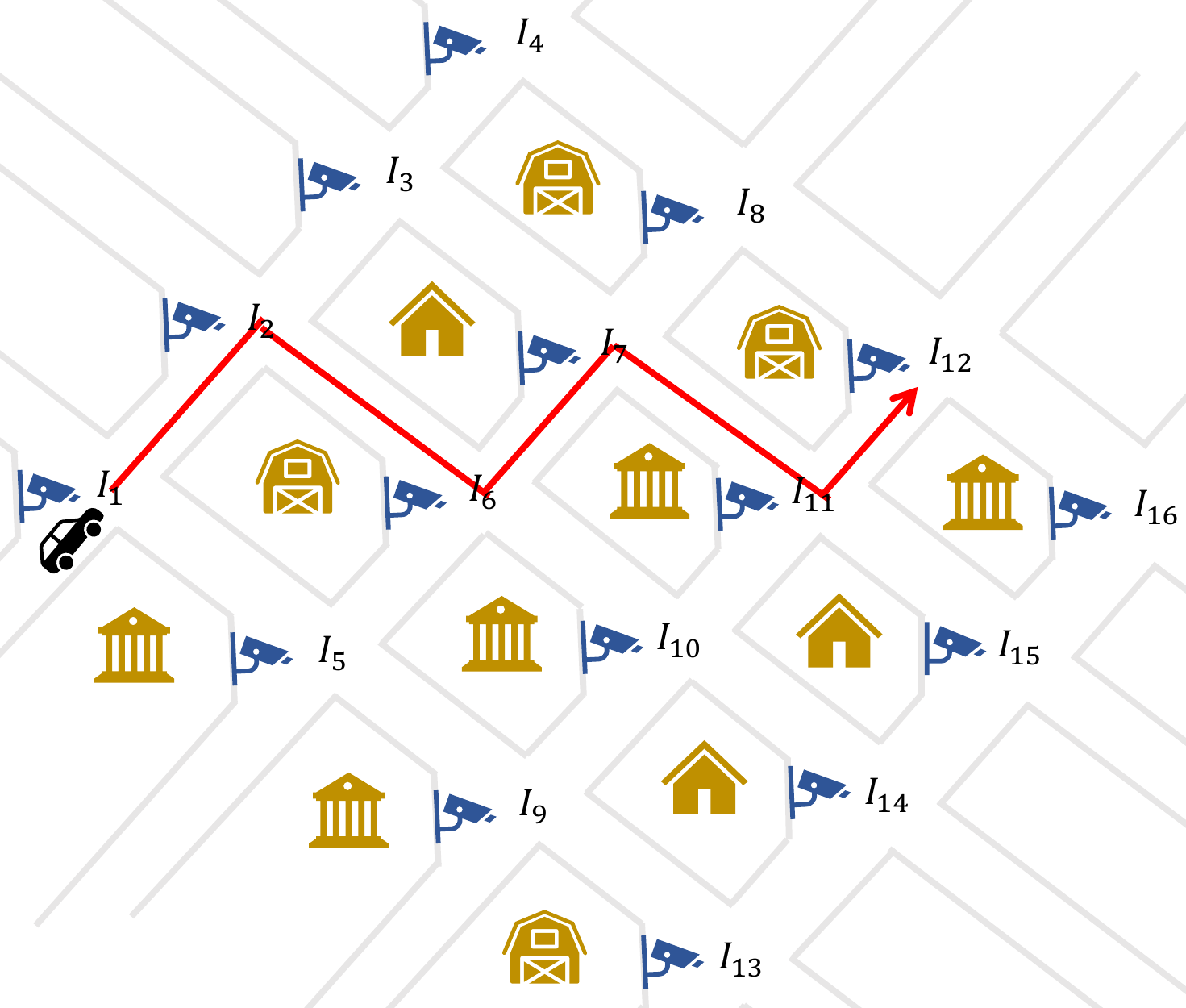}
\caption{\small An example of vehicle real-time tracking}
\label{fig_scinario}
\vspace{-15pt}
\end{figure}

%\begin{figure}[t!]
%\centering
%\includegraphics[width=2in]{Figs/graph.pdf}
%\caption{Graph for the road network.}
%\label{fig_graph}
%\end{figure}

\subsection{Tracking Loss Issue}
Fig.~\ref{fig_scinario} shows an example of a road network consisting of 16 intersections. Suppose a hit-and-run vehicle has a trajectory of $T = \{I_{1} (t_1, t_2),I_{2} (t_3, t_4),I_{6} (t_5, t_6), I_{7} (t_7, t_{8}), I_{11} (t_{9},t_{10}),$ $I_{12} (t_{11},t_{12})\}$ in the monitored area. Then, the edge nodes deployed at $I_{1}, I_{2}, I_{6}, I_{7}, I_{11}, I_{12}$ are involved in tracking the VoI. If we ignore the time periods taken by the vehicle at the intersections, the trajectory $T$ can be considered as a path in the road network, i.e., $I_1\rightarrow I_2\rightarrow I_{6}\rightarrow I_{7}\rightarrow I_{11}\rightarrow I_{12}$. In this paper, the tracking system aims to track the VoI in real-time: if the VoI enters $I_i$ at time instant $t_p$, then 

\begin{itemize}
    \item the edge node at $I_i$ must be activated before $t_p$. Otherwise, the VoI may not be captured by the camera and get lost at $I_i$.
    \item the VoI and the next intersection $I_j$ on the VoI's trajectory must be identified before the time instant $t_q$ when the VoI enters $I_j$. Otherwise, the tracking system cannot activate the edge node at $I_j$ before $t_q$. 
\end{itemize}

\begin{example}
For the example in Fig.~\ref{fig_scinario}, a Mercedes silver GLB SUV (the VoI) is reported to be involved in a hit-and-run accident at intersection $I_1$ at time instant $t_1$ and it arrives at $I_2$ at $t_3$. The edge node at $I_1$ is activated at $t_1$: the camera captures videos from $I_1$ and the videos are processed frame-by-frame to identify the VoI and the next intersection on the VoI's trajectory. The processing results must be obtained and sent to the cloud before $t_3$. Then, the tracking system can activate the edge node at $I_2$ before $t_3$; otherwise, the VoI may already depart from $I_2$ before the edge node starts tracking.
\end{example}

Note that the next intersection on the VoI's trajectory can be easily determined by the VoI's position and the road network based on a series of video frame processing.  

If the tracking system tracks the VoI in real-time successfully, the VoI will be located at either an intersection or a road segment between two intersections at any specific time instant when it travels in the monitored area. The tracking problem becomes rather trivial to resolve if only a few vehicles travel in the smart city. It means that whenever the VoI appears at an intersection, it will be recognized by the object identification algorithms instantly due to sufficient computing resources on the edge node. However, the practical scenario is always not the case.

\begin{example}\label{example:loss}
In Fig.~\ref{fig_scinario}, the VoI arrives at intersection $I_2$ at time instant $t_3$ and travels to $I_6$. Assume that $I_2$ is a crowded intersection. In this case, it takes a very long time to identify the VoI from dozens of cars traveling through the intersection. The edge node at $I_2$ may fail to complete the video analytics when the VoI arrives at $I_6$ at $t_5$. Then, the tracking system does not know which edge nodes should be activated afterward to carry on the tracking task and the VoI is lost. 
\end{example}

\noindent{\bf Identified Issue.} The computing capacity of an edge node depends on the computing platform used to implement it, which is determined by the hardware manufacturer. For example, an Azure Data Box Edge~\cite{Azure} node is equipped with $2 \times 10$ core CPUs for data processing. Thus, at the crowded intersections, the tracking system cannot identify the VoI in real-time due to limited computing capacity of the edge node and the ``tracking loss'' occurs.

\noindent{\bf Key idea of our proposed method.} To resolve this ``tracking loss'' issue, we seek to develop a smart tracking method to make up for the limited computing capacity of a single edge node. At a crowded intersection, since the full-fledged object identification algorithm needs an unaffordable amount of time to precisely find the VoI, the tracking system can use a ``coarse'' object identification algorithm to save time. Multiple suspected VoIs may be identified by the ``coarse'' algorithm and travel to different intersections. The tracking system can track all the suspected VoIs simultaneously by utilizing the edge nodes at different intersections and identify the VoI at the uncrowded intersections. Intuitively, this idea is feasible, because we find that almost 95\% of the intersections in a smart city are uncrowded intersections (See the statistics in Sec.~\ref{sec:traffic}).

In order to implement the real-time tracking system, first we propose a dynamic Vehicle Re-Identification (Re-ID) framework to realize the Re-ID algorithm at different granularity levels. Then we introduce a real-time admission control module on each edge node to decide which Re-ID algorithm will be performed to identify the VoI according to the number of vehicles detected at the corresponding intersection. Finally, we will discuss how to obtain the active period for each edge node involved in a hit-and-run tracking event and how the proposed tracking system, named \emph{WatchDog}, works to track the VoI in real-time.
%!TEX root = main.tex

\section{Vision-based Vehicle Tracking}

Tracking in \name\ relies on computer vision-based machine learning algorithms. Query input is a target vehicle (e.g., from an accident report) with information such as make, model, color, and plate number. Once \name\ receives a tracking query, the corresponding edge node starts running a video analytics pipeline to analyze traffic videos in real-time. At a high level, the processing pipeline of each frame consists of two modules: vehicle detection and vehicle re-identification. 

\subsection{Vehicle Detection}
\label{sec:vd}
For each video frame, vehicle detection targets to find all vehicles and assign class to each one of them. %Thus, vehicle detection has two tasks: vehicle localization and vehicle classification. 
Unlike classification networks, traditional vehicle detection networks \cite{ross2015RCNN, ren2015FRCNN, dai2016RFCNN, he2017maskrcnn} have three components: CNN-based feature extractor, Region Proposal Network (RPN) and a classifier. They usually use a pre-trained classification network (e.g., ResNet) as feature extractor to a generate feature map of an input image. After that, they utilize RPN \cite{ross2015RCNN} to generate all candidate bounding boxes of vehicles, and finally assign labels for each bounding box. As those detectors extract all vehicles first and classify each bounding box later, they are called two-stage vehicle detectors. Although two-stage vehicle detectors achieve superior performance on many public benchmarks \cite{boxy2019, marius2016cityscapes, geiger2012CVPR}, the speed suffers. %In many embedding systems, latency is a very important component for usability. 
To trade-off performance and latency, researchers seek to design efficient vehicle detectors \cite{redmon2016yolov1, liu2016ssd, lin2017RetinaNet, li2018yolov3} which use one CNN network to solve localization and classification simultaneously. %Because one CNN network have limited ability to extract powerful feature maps, their performance is typically weaker than two-stage vehicle detectors. These efficient vehicle detectors are also called one-stage vehicle detection methods. 
Albeit marginal performance drop, one-stage detectors largely reduce latency, and hence have been widely implemented on today's edge nodes for real-time tracking system.

\subsection{Vehicle Re-identification}
\label{sec:vd2}
Vehicle re-identification (V-ReID) determines whether two bounding boxes belong to the same vehicle. The most popular deep learning approach for V-ReID is to build a CNN-based feature extractor and differentiate bounding boxes based on the similarity (e.g., cosine distance) between discriminative feature vectors. Unlike person re-identification (P-ReID), V-ReID \cite{khorramshahi2019vreid, tan2019vreid, huang2019vreid, spanhel2019vreid, hsu2019vreid} often uses many CNN-based networks to extract different features (e.g., global features, region features, key point features) and concatenate them for a reliable comparison. For example, researchers often set features of vehicle's shape as a general feature and window screen as a region feature. As a result, V-ReID models are often very large and the end-to-end compute process is prohibitively costly, making them not amenable to real-time tasks. For example, V-ReID each one of a large set of vehicle bounding boxes during the rush hours could end up causing ``tracking loss'' issue. Thus, a V-ReID method that can dynamically trade-off accuracy with inference time in real-time is desired. To this end, in what follows we propose a dynamic V-ReID framework called D-V-ReID.

% Fortunately, an accurate V-ReID may be replaced by a sequence of the detection processes \cite{oliveira2019vreid, liu2018vreid} from the coarse methods (e.g., color and model matching) to the find-grained ones (e.g., licenses plate matching).

\subsection{D-V-ReID Pipeline}
\label{sec:sub}
D-V-ReID adopts the idea of divisive clustering where V-ReID on each frame follows a multi-layer framework where upper layers correspond to coarse but efficient classifications and lower layers are in charge of powerful but complex detection. While we need to slightly modify the existing learning algorithm for the purpose of real-time tracking, we do not intend to propose any new learning algorithm in this paper but focus on building a flexible V-ReID pipeline. The cascaded pipeline we propose includes: color matching, model and make matching, and the full-fledged deep feature-based re-identification.  

\subsubsection{Color Matching} As a basic filter, color matching measures similarity between bounding boxes by color distribution in spaces such as RGB or HSI \cite{gonzales2002digital}. Given RGB or HSI features, we use K-Nearest Neighbors (KNN) \cite{guru2010knn} to decide the similarity of two images. A car is classified by a majority vote of its neighbors, with the car being assigned to the class which is most common amongst its $k$ nearest neighbors. The complexity of the inference process in this step is linear to the input frame size, and it can be further reduced by frame down-sizing, which provides a trade-off between matching quality and inference time where the inference time can be made arbitrarily small. As the most coarse V-ReID step, a matching confirm result does not provide much confidence that the target vehicle is detected because different vehicles may have similar color, but a matching reject serves as a strong evidence for the fact that the target vehicle is not in the current frame. 

\begin{figure}[t!]
\centering
\includegraphics[width=3.25in]{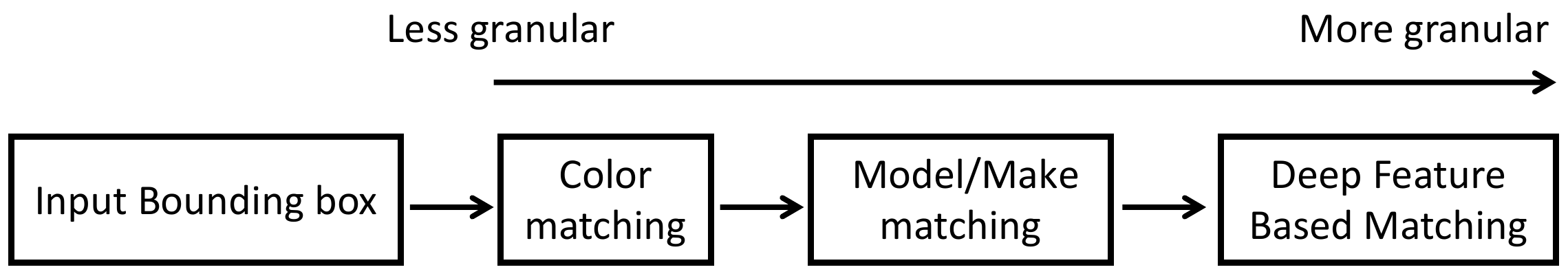}
\caption{\small Layered modules.}
\label{fig: framework}
\vspace{-15pt}
\end{figure}

\subsubsection{Model and Make Matching} Our second layer aims to detect if the vehicle in the bounding box has the same model and make as the target vehicle. Model and make of a vehicle are more distinctive, and such information is commonly used for vehicle identification in security systems such as amber alert. Traditional matching algorithms use Scale Invariant Feature Transform (SIFT) \cite{david1999SIFT}, SURF \cite{herbert2008surf} and HOG \cite{dalal2005hog} feature extraction techniques to extract models' features. Based on these features, Support Vector Machine \cite{baran2015SVM,noppakun2017svm} and Random Forest \cite{galiano2012random} can matching two cars accurately. But these methods assume that the input image has the whole shape of vehicle. In practical environments, many cars are occluded by other objects. To resolve this issue, many work \cite{faezeh2017deep, jo2018deep} seeks to use deep learning approaches to generate discriminative features because of the ability to extract features automatically. Unlike general image classification or object detection, model and make matching on vehicle is done by training a light convolutional neural network (ResNet50, MobileNet, ShuffleNet, EfficientNet \cite{he2016deep, sandler2018mobilenetv2, zhang2018shufflenet, tan2019efficientnet}). For instance, we could adopt a subset of ResNet for such purpose. The goal of model and make matching in our framework is still to provide a fairly accurate V-ReID method with moderate inference overhead. %Note that we do not use landmark matching in this step, and the model/make matching is done according to the appearance. %Finally, the color matching in the last step will be utilized as a submodule.

%The start-of-the-art V-ReID methods are designed through convolutional neural networks, and these methods often involve several learning blocks combined together for a unified loss function. In particular, we first training a deep metric embedding  for extracting hidden features as well as one attribute extractor for identifying two hand-craft features: pose and landmarks \cite{nguyen2019vehicle}. These two modules will be fused together followed by multiplayer perceptrons.  Different from most of the existing methods, we assign each sub-model a loss function and a output gate, and the whole network is trained as a multi-task learning. The reason of doing so is to provide early exits such that the V-ReID in this part can be further decomposed. The results in the previous two steps, color matching and model/maker matching, are used as input features. 

\subsubsection{Full-fledged V-ReID}
\label{deep_feature}
Deep feature-based V-ReID algorithm serves as the final layer in D-V-ReID. In light of ensemble learning \cite{david1999ensemble}, start-of-the-art V-ReID methods \cite{huang2019vreid, pirazh2019vreid, he2019vreid, wang2017vreid, georgia2019vreid} are designed with a set of convolutional neural networks, and these neural networks are responsible for different feature extractions. To further improve ReID accuracy, researchers assign unique lose functions \cite{tan2019vreid} to different feature extractors and train them independently. After pre-training, a unified loss function is utilized to train all feature extractors together. This step is also called multi-task learning \cite{zhang2018multitask, chen2017multitask, huang2019vreid, georgia2019vreid}. In particular, these methods always classify features as global, region and key-point features. Global feature is used to describe the overall appearance of the vehicle, which is the linear transformation of the pooling feature of the last convolution module in global feature extractor. Due to the limited discriminative ability of global features, many researchers seek to use multiple granularities network \cite{wang2018granularities} to extract features from the multiple semantic parts from vehicles. %However, many detailed information that is hard to be found by global features and semantic part features, such as logo, light, etc. 
Inspired by \cite{wang2017vreid}, scientists use another CNN network to predict several key points sit on key parts of the vehicle, and then extract feature around those key parts with the assistance of a heatmap generated from key point network. In our design, we combine features from color matching and model/make  matching, and set it as the global feature. The reason for doing so is to provide early exit and reuse computation from previous matchings.

%\subsubsection{License Plate Matching} The last level of our framework aims to detect if the license plate of the vehicle in the bounding box is identical to that of the target vehicle. An true-matching of license plate would be a very strong evidence of identification. This step is particular challenging as the appearance of the license plate depends on illumination, angle, and the letter structure specified by the state. We will train one model for point of interested (POI) locating the plate, and then perform matching through a convectional neural network with LSTM for sequence analysis \cite{li2016reading}. For our problem, it offers a little solace that the angle is fixed at a certain intersection, meaning that the location of the plate is relatively fixed to the bounding box. \ys{We can remove this module if we do not have a handy license plate recognition library. }

\subsection{D-V-ReID framework.}
D-V-ReID framework is built upon three V-ReID algorithms with different granularity levels, which is illustrated in Fig.~\ref{fig: framework}. As discussed above, we set K-Nearest Neighbors classifier \cite{guru2010knn} as our color filter and the architecture of model matching to MobileNetV2 \cite{sandler2018mobilenetv2}. In terms of the full-blown V-ReID, we adopt the best method \cite{tan2019vreid} in AI City Challenge 2019. They utilized three different convolution neural networks to extract features from the same vehicle and concatenate them as the final feature. As discussed in Section~\ref{deep_feature}, we combine features from color matching and model/make matching as the global feature. To see the computation cost of region and key-point feature extractors, we set their architectures to ResNet101 \cite{he2016deep} and SE-ResNet152 \cite{hu2018senet}, respectively. After re-scaling each bounding box to $224*224$, we report floating point operations per second (flops), inference time and cpu usage on one bounding box of different feature extractors from deep feature based matching in Table~\ref{cost_deep} and different modules in Table~\ref{cost}. The test environment is AMD Ryzen 7 3700x (CPU) with 4G memory. Because K-Nearest Neighbors classifier is not a deep neural network, we don't record the flops for it.

\begin{table}
    \caption{Cost of deep feature extractors (global, region and key-point features).}
    \begin{center}
    \begin{tabular}{|c|c|c|c|}
    \hline
    %\textbf{PC}&\multicolumn{5}{|c|}{\textbf{Contribution Scores}} \\
    %\cline{2-6} 
     & \textbf{Global} & \textbf{Region}& \textbf{Key-point} \\
    \hline
    \hline
    Flops & 341.5M & 7868.4M & 11392.M \\
    \hline
    Inference Time & 41.1ms & 96.7ms & 172.3ms \\
    \hline
    CPU Usage & 2.4\% & 9.8\% & 15.0\% \\
    \hline
    \end{tabular}
    \label{cost_deep}
    \end{center}\vspace{-15pt}
\end{table}

\begin{table}[h]
    \caption{Cost of different modules (color matching, model matching and full-fledged V-ReID).}
    \begin{center}
    \begin{tabular}{|c|c|c|c|}
    \hline
    %\textbf{PC}&\multicolumn{5}{|c|}{\textbf{Contribution Scores}} \\
    %\cline{2-6} 
    & \textbf{Color} & \textbf{Model}& \textbf{Full-fledged V-ReID} \\
    \hline
    \hline
    Flops & $/$ & 341.5M & 19602.3M \\
    \hline
    Inference Time & 0.5ms & 40.6ms & 310.1ms \\
    \hline
    CPU Usage & 0.3\% & 2.2\% & 27.3\% \\
    \hline
    \end{tabular}
    \label{cost}
    \end{center}\vspace{-10pt}
\end{table}

According to the real-time requirements of the tracking system, it could select the V-ReID modules with a proper complexity to identify the VoI at specific intersections. Intuitively, in order to guarantee that the V-ReID module can complete in real-time, a less granular module will be triggered if a large number of vehicles appear at a crowded intersection. On the contrary, if the intersection is uncrowded, a more granular module will be performed to identify VoI. Note that under our proposed D-V-ReID framework, if a more granular module is selected, all the less granular modules are executed implicitly. It is evident that the granularity selection depends on the number of vehicles captured on each video frame, the inference time of different V-ReID modules and the computing capacity of the edge node. In next section, we introduce a real-time admission control method, which is implemented on each edge node to select proper V-ReID modules for the tracking system in real-time.

%Note that we can continuous the matching process regardless the results of the upper layers. For example, even if deep feature based matching has already rejected the input, we could still use license plate matching to further confirm the matching result. \ys{What's the point of doing so?} The training process can leverage either the benchmarks such as \cite{yang2015large, liu2016large, liu2016deep}, or the newly correct data at each intersection. Using benchmarks \ys{benchmark? pre-trained models?} could reduce the training burden as each intersection would utilize the same model for inference, while training each box edge with local data could produce a better V-ReID result. Once the exit has been selected, the inference time at each intersection is determined. Thus, an trade-off between the quality of V-ReID and the inference time is established. Given the response time requirement, we would be able to select an appropriate exit. Finally, a fine-grained trade-off can be found by using sub-frame sampling in color matching as well as branch skipping, for example, from model/maker matching to license plate matching without the deep feature based matching.
%!TEX root = main.tex

\section{Real-time Admission Control}

Before discussing the details of our design, we introduce some preliminary results on real-time admission control, which determines the granularity level of the V-ReID modules selected to identify the VoI.

\subsection{Real-Time Task Scheduling Framework}
In this section, we will introduce a classic soft real-time schedulability test, which can be used to calculate the completion time bounds for real-time tasks scheduled in the real-time system. Based on the completion time bound of each real-time task, we can perform admission control on each edge node.

\subsubsection{Real-time Task Model} 

At an arbitrary intersection $I_x$, the surveillance camera captures frames from the intersection periodically and the \emph{period}, denoted by $p$, depends on the camera's frequency. For example, if the camera captures 24 frames every second, we would say the video is 24 fps and its period is $\frac{1}{24}s$. Usually, one or more vehicles may be detected on each frame. The V-ReID machine learning algorithm is performed on all the detected vehicles to identify the VoI and each identification process corresponds to \emph{a real-time machine learning task}. Let $e_i$ denote the \emph{processing time} of task $\tau_i$ performed on vehicle $i$. Tabel~\ref{cost} gives the processing time for different V-ReID modules. Let $e^c, e^m, e^d$ denote the processing time for color matching module, model/make matching module and Deep Feature Based Matching module, respectively. Under our proposed D-V-ReID pipeline framework, if a more granular module is selected, all the less granular modules has already been selected implicitly. Thus, $e_i$ has three different options under the D-V-ReID framework: $e^1_i = e^c, e^2_i = e^1_i + e^m, e^3 =  e^2_i + e^d$. As we discussed in Sec.~\ref{sec:system}, in order to avoid ``tracking loss'', the VoI must be identified before it arrives at the next intersection (it is illustrated by Example~\ref{example:loss}). 

%the VoI must be identified before $t_7$ when it arrives at $I_8$. In other words, $t_7$ is \emph{a deadline} for all the real-time identification tasks performed at $I_7$ and the real-time tasks must complete by $t_7$. If anyone of the tasks misses the deadline, ``tacking loss'' may occur. 

\begin{definition}\label{def:deadline}
 Let $t_{x,y}$ denote the travelling time of the VoI between two neighbouring intersections (i.e., $I_x$ and $I_y$). $I_x$ may have multiple neighbouring intersections and we define the shortest $t_{x,y}$, denoted by $D_x$, as the \emph{relative deadline} of the real-time tasks from the edge node deployed at $I_x$. 
\end{definition}

Based on the above discussion, we use the periodic hard real-time task model to describe the execution behaviors of real-time workloads in the tracking system on an edge node deployed at $I_x$. We consider the problem of scheduling $n$ periodic real-time tasks on $M$ processors. That means $n$ vehicles are detected from each image and the edge node has $M$ processors (note that we use ``processor'' to denote the minimum schedulable processing unit). A task $\tau_i$ is characterized by two parameters - a processing requirement $e_i$ and a period $p$ with the interpretation that the task generates a job (i.e., the camera captures a frame) in every $p$ time units and each such job $\tau_{i,j}$ has a processing requirement of $e_i$ execution units which should be met by a deadline $d_{i,j}$. Let $r_{i,j}$ denote the generation time of $\tau_{i,j}$, then $d_{i,j} = r_{i,j} + D_x$. We further let $u_i$ denote the utilization of $\tau_i$, where $u_i = \frac{e_i}{p}$, and the utilization of the task system $\tau$ is defined as $U_{sum} = \sum_{i=1}^nu_i$. Successive jobs of the same task are required to execute in sequence. We require $u_i \leq 1$, and $U_{sum} \leq M$; otherwise, deadlines will be missed.

\subsubsection{Real-time Scheduling Algorithm}

If the number of tasks is no greater than the number of processors, each identification task can be executed on a dedicated processor. In this case, the computing capacity on the edge node is sufficient to execute the real-time tasks and the ``tracking loss'' issue will not happen. However, when the traffic is heavy, the number of vehicles (corresponding to the real-time tasks) detected at the intersection may be much larger than the number of processors on the edge node. The real-time tasks will compete for the limited computing resources on the edge node and have inferences with each other, which may give rise to a huge delay for frame processing and lead to deadline miss. In this case, a scheduling algorithm is needed to allocate processor time to tasks, i.e., determines the execution-time intervals and processors for each job while taking any restrictions, such as on concurrency, into account. In real-time systems, processor-allocation strategies are driven by the need to meet timing constraints and in our real-time tracking system, we apply the First-in First-Out (FIFO) policy to schedule the real-time tasks: processors execute jobs in the exact order of job arrival.

\subsubsection{Completion Time Analysis}

A given set of real-time tasks is said to be schedulable on a given system of processors if the tasks can be scheduled on these processors in such a manner that all jobs of all the tasks always complete by their deadlines. Physically, if all the tasks can complete by their deadlines, the ``tracking loss'' will not happen. The schedulability can be verified by using standard schedulability analysis (Theorem~\ref{theorem:test}) for calculating the completion times of real-time tasks.

%Thus, we can use the classic schedulability analysis (Theorem~\ref{theorem:test}) for FIFO scheduler to validate the schedulability of real-time tasks scheduled in our tracking system.

\begin{theorem}\label{theorem:test}
Considering that a real-time task system $\tau$ of $n$ periodic tasks are scheduled on $M$ processors under FIFO scheduling policy, the completion time bound for a task $\tau_i$ is
\vspace{-5pt}
\begin{equation}
R_i = p + e_i + \frac{\sum_{\tau_k\in E(\tau, M-1)}e_k - e_i}{M - \sum_{\tau_j\in U(\tau, M-1)}u_j}
\end{equation}
where $E(\tau, M-1)$ denotes the set of at most (M-1) tasks with the
highest execution costs from $\tau$ and $U(\tau, M-1)$ denotes the
set of at most (M-1) tasks of highest utilization from the task
set $\tau$~\cite{leontyev2007tardiness} . 
\end{theorem}
\vspace{-5pt}

The proof of Theorem~\ref{theorem:test} is given in~\cite{leontyev2007tardiness} and the same result can be derived from~\cite{dong2018general}, because the periodic task model is a special case of the stochastic task model discussed in~\cite{dong2018general}. In light of the real-time task model and the completion time analysis, a formal definition of ``tracking loss'' is that if the completion time bounds of the real-time tracking tasks exceed their deadlines, the VoI is lost at some intersections.

%\begin{example}
%In Example~\ref{example:loss}, when the VoI arrives at $I_7$, the camera deployed at $I_7$ periodically captures frames from $I_7$. Since the traffic at $I_7$ is very heavy, a lot of vehicles are detected in this intersection. That means a large number of machine learning tasks are generated on the edge node to identify the VoI among the detected vehicles. Tasks compete for computing resources on the computing platform with a limited number of processors and the executions of some tasks are delayed, which could result in a deadline miss. The tracking tasks cannot complete before $t_7$ and the VoI is lost.
%\end{example}

\subsubsection{Real-time Admission Control}
Intuitively, if we can guarantee that the completion time bound $R_i$ for each task $\tau_i$ can be no greater than its deadline $D_x$, the ``tracking loss'' cannot happen. According to Theorem~\ref{theorem:test}, since $p$ and $M$ are fixed values when the hardware of the edge node is given, the completion time bound $R_i$ of $\tau_i$ only depends on the execution times of the machine learning tasks and the number of vehicles detected at the intersection. Therefore, we introduce the following programming to select proper V-ReID modules for the tracking system in real-time.
\vspace{-5pt}
\begin{equation}
\begin{array}{ll@{}ll}
\text{maximize}  & \displaystyle\sum\limits_{i=1}^{n} e_i, &\\
\text{subject to}& R_i = p + e_i + \frac{\sum_{\tau_k\in E(\tau, M-1)}e_k - e_i}{M - \sum_{\tau_j\in U(\tau, M-1)}u_j} &\leq D_x   ,  &\\
                 &  e_i \in \{e_i^j, e_i^{j+1}\}, &j = 1, 2
                
\end{array}\label{eq:programming}
\end{equation}

\noindent where $e_i$ denotes the execution time of task $i$ in the $j^{th}$ iteration. According to Tabel~\ref{cost}, the processing time of color matching module is $0.5~ms$. We assume that the edge node has enough computing capacity to perform color matching module on all detected vehicles even at the most crowded intersection. Thus, in the first iteration, our objective is to maximize the number of tasks which will perform the model/make matching module. If all the tasks can meet the deadline, then in the second iteration, we aim to maximize the number of tasks which will perform the deep feature matching module. $R_i$ is the response time of task $i$, which is defined in Theorem~\ref{theorem:test}, and $D_x$ is the relative deadline of every real-time task from $I_x$, which is defined by Def.~\ref{def:absdeadline}. We require that each task must complete by its deadline. Since we do not have preferences for the tasks which will perform in the $j^{th}$ iteration, the time complexity for this programming is $O(n)$ to achieve the optimal solution.
%!TEX root = main.tex

\section{Active Periods of Edge Nodes}\label{sec:active}

In order to track the VoI in real-time, the tracking system should be able to know when the VoI will arrive at which intersections in advance, then the corresponding edge nodes can be activated before the VoI's arrival time and identify the VoI when it appears. In other words, the active period of an involved edge node should cover the time interval when the VoI travels though the corresponding intersection. 

If all the intersections are uncrowded, this problem is trivial. For the example in Fig.~\ref{fig_scinario}, when the hit-and-run accident is reported, the edge node at intersection $I_2$ is activated to perform the most granular machine learning algorithm on all detected vehicles and the VoI is identified at time instant $t_1$. Based on the real-time video analytics, the tracking system finds that the VoI departs from $I_2$ and travels to $I_3$ at time instant $t_2$. Then, the tracking system activates the edge node at $I_3$ instantly and at the same time stops the edge node at $I_2$. If the tracking system repeats the same operation on all involved edge nodes, the active period for each edge node can be obtained in real-time.

\begin{figure}[t]
\centering
\subfloat[Travelling time at an $I_x$.]{%
\includegraphics[width=9.6pc]{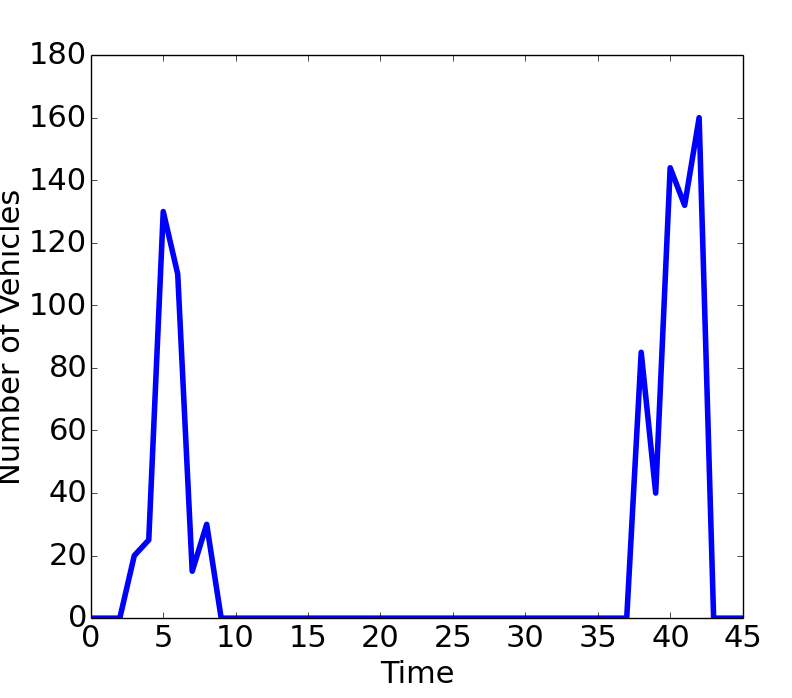}}
\label{fig:intersection}\hfill
\subfloat[Travelling time on a road segment.]{%
\includegraphics[width=11.2pc]{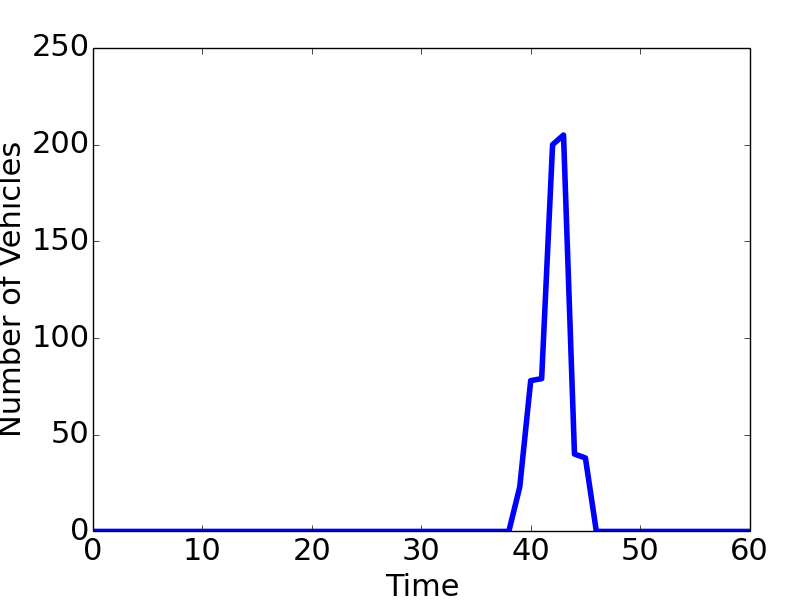}}
\label{fig:segment}\hfill
\caption{\small Travelling time measurement at different locations. The x-axis denotes the travelling time and y-axis denotes the number of vehicles travelling through this location.}
\label{fig:travellingtime}
\vspace{-15pt}
\end{figure}

However, when the VoI enters a crowded intersection, the problem becomes challenging. Assume that intersection $I_3$ is a crowded intersection in Fig.~\ref{fig_scinario} and the VoI (a Mercedes silver GLB SUV) is traveling from $I_2$ to $I_3$. The admission control module selects the color matching module to track all the silver vehicles travelling through $I_3$ based on the number of vehicles detected in the video frames. Multiple silver vehicles may appear in $I_3$, but the tracking system cannot confirm that whether or not the VoI has arrived at $I_3$. Because the travelling time on the road segment between $I_2$ and $I_3$ is different for different vehicles. Fig.~\ref{fig:travellingtime} shows the travelling time measurement at different locations. On a road segment, the travelling time of vehicles varies within a range from 30 seconds to 50 seconds (Fig.~\ref{fig:travellingtime}.b). Thus, at a crowded intersection, in order to guarantee that VoI is included in the tracked vehicles (in the example, VoI is one of the tracked silver vehicles), the active period for the edge node is calculated based on historical traffic information.

\begin{definition}\label{intersection_time}
Let $t_x$ denote the travelling time taken by the VoI at intersection $I_x$, then $t_x^l\leq t_x\leq t^u_x$, where $t_x^l$ is the lowest value among all the vehicles' travelling time at $I_x$ and $t^u_x$ is the largest value. Both values can be obtained from historical traffic information (For example, Fig.~\ref{fig:travellingtime}.a).  
\end{definition}

\begin{definition}\label{segment_time}
Suppose $I_x$ and $I_y$ are neighbouring intersections. Let $t_{x,y}$ denote the travelling time taken by the VoI at road segment $e_{x,y}$, then $t_{x,y}^l\leq t_{x,y}\leq t^u_{x,y}$, where $t_{x,y}^l$ is the lowest value among all the vehicles' travelling time at $e_{x,y}$ and $t_{x,y}^u$ is the largest value. Both values can be obtained from historical traffic information (For example, Fig.~\ref{fig:travellingtime}.b).   
\end{definition}

\begin{definition}\label{def:absdeadline}
Note that according to Definition~\ref{def:deadline} and Definition~\ref{segment_time}, $I_x$ may have multiple neighbour intersections and we use the shortest $t^{x,y}_l$ as the \emph{relative deadline} $D_x$ of the real-time tasks from the edge node deployed at $I_x$. 
\end{definition}

Based on Definition~\ref{intersection_time}, Definition~\ref{segment_time} and Definition~\ref{def:absdeadline}, we can calculate the active period $[t_x^s, t_x^e]$ for an involved edge node at intersection $I_x$.
\begin{itemize}
    \item {\bf Case 1:} If the previous intersection $I_p$ on the VoI's trajectory is an uncrowded intersection, then the VoI is identified at $I_p$. Let $t_p$ denote its departure time from $I_p$. Then, the earliest time instant when the VoI can arrive at $I_x$ is $t_x^s = t_p + t_{p,x}^l$. Correspondingly, $t'_x = t_p + t_{p,x}^u$ is the latest time instant when the VoI can arrive at $I_x$. Thus, the latest time instant when the VoI departs from $I_x$ is $t'_x + t_x^u$, which is the time instant when the last video frame captured from $I_x$ may contain the VoI. Thus, the edge node needs $D_x$ time units to process the last frame and the edge node ends up processing video frames at $t_x^e =  t'_x + t_x^u + D_x$.
    
    \item {\bf Case 2:} If the previous intersection $I_p$ is a crowded intersection and we assume that the active period of the edge node at $I_p$ is $[t_p^s, t_p^e]$. According to the definition of active period, the earliest time when the VoI enters $I_p$ is no earlier than $t_p^s$, then the earliest time when the VoI enters $I_x$ is no earlier than $t_x^s = t_p^s + t_x^l + t_{p,x}^l$; the latest time when the VoI departs from $I_p$ is no later than $t_p^e - D_p$, then the latest time when the VoI departs from $I_x$ is no later than $t_p^e - D_p + t_x^u + t_{p,x}^u$. Again, the edge node needs $D_x$ time units to process the last frame and the edge node ends up processing video frames at $t_x^e =  t_p^e - D_p + t_x^u + t_{p,x}^u + D_x$.
\end{itemize}

In light of the above discussion, the active period for the edge node at intersection $I_x$ is calculated based on the active period of the previous edge node. Thus, we need to define the active period for the first edge node at intersection $I_o$ where the hit-and-run accident happens. Let $t_o$ denote the time instant when the accident is reported. Then, whether or not $I_o$ is a crowded intersection, the active period of the edge node at $I_o$ is $[t_o, t_o + t_o^u + D_o]$. Based on $I_o$'s active period, the active periods for all the involved edge nodes can be calculated one-by-one at run time.  

\section{WatchDog}\label{sec:watchdog}
In this section, we put all the proposed techniques together to build the real-time tracking system - WatchDog.

\begin{figure}[t!]
\centering
\includegraphics[width=3.5in]{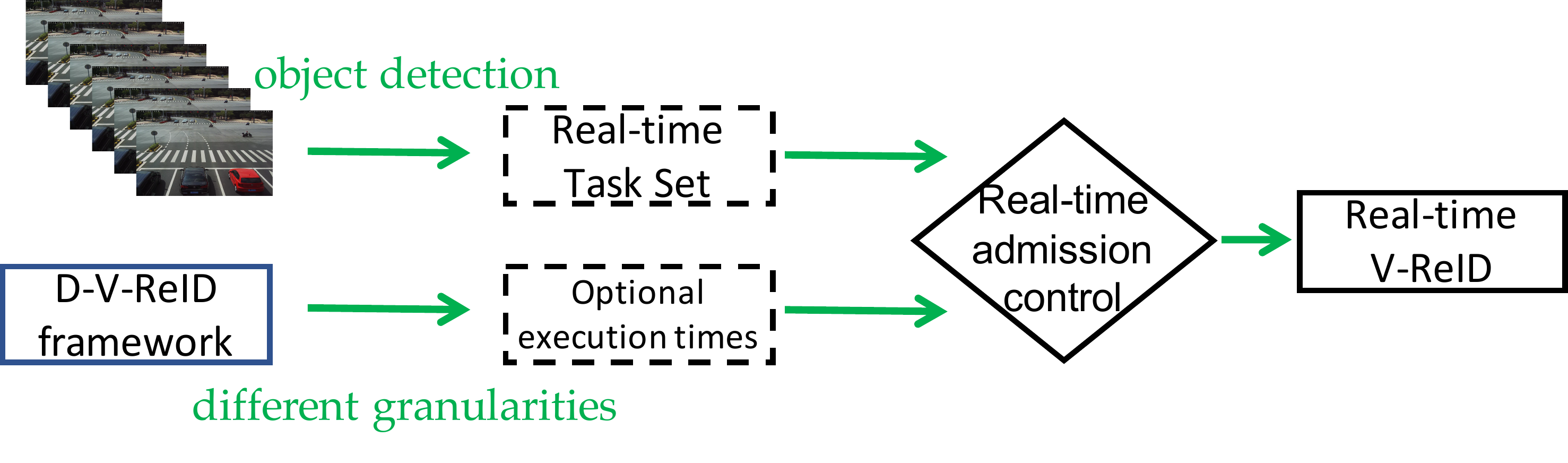}
\caption{\small The architecture of WatchDog implemented on the edge node.}
\label{fig:watchdog}
\vspace{-15pt}
\end{figure}

\subsection{System Description}\label{sec:system}
Fig~\ref{fig:watchdog} illustrates the architecture of WatchDog implemented on each edge node, which consists of four major components: (\textit{i}) a live video stream generated by the surveillance camera; (\textit{ii}) D-V-ReID framework; (\textit{iii}) the Real-time admission control module; (\textit{iv}) the Real-time Vehicle ReID program. On each edge node, each frame of the real-time video stream is firstly processed by a vehicle detection module to detect the vehicles traveling through this intersection, which is introduced in Sec.~\ref{sec:vd}. A real-time ReID task will be performed on each detected vehicle to identify whether or not the VoI appears. Suppose the system detected $n$ vehicles at this intersection, then the real-time task set contains $n$ tasks. According to the D-V-ReID framework introduced in Sec.~\ref{sec:vd2}, each task has several optional execution times, which correspond to different sub-modules, which is introduced in Sec.~\ref{sec:sub}. The longer a real-time task executes, the better ReID performance the real-time task can get. In order to guarantee that each real-time task can complete by its deadline, the real-time admission control module is performed to select the best combination of execution times for real-time tasks to ReID the VoI, according to the optimal solution for the programming, which is given by Eq.~\ref{eq:programming}. Intuitively, it is a trade-off between the ReID performance and the schedulability of the real-time task system. When the modules of ReID program are chosen for each real-time task, the real-time V-ReID starts performing on each frame to track the VoI.

\begin{figure}[t!]
\centering
\includegraphics[width=2.5in]{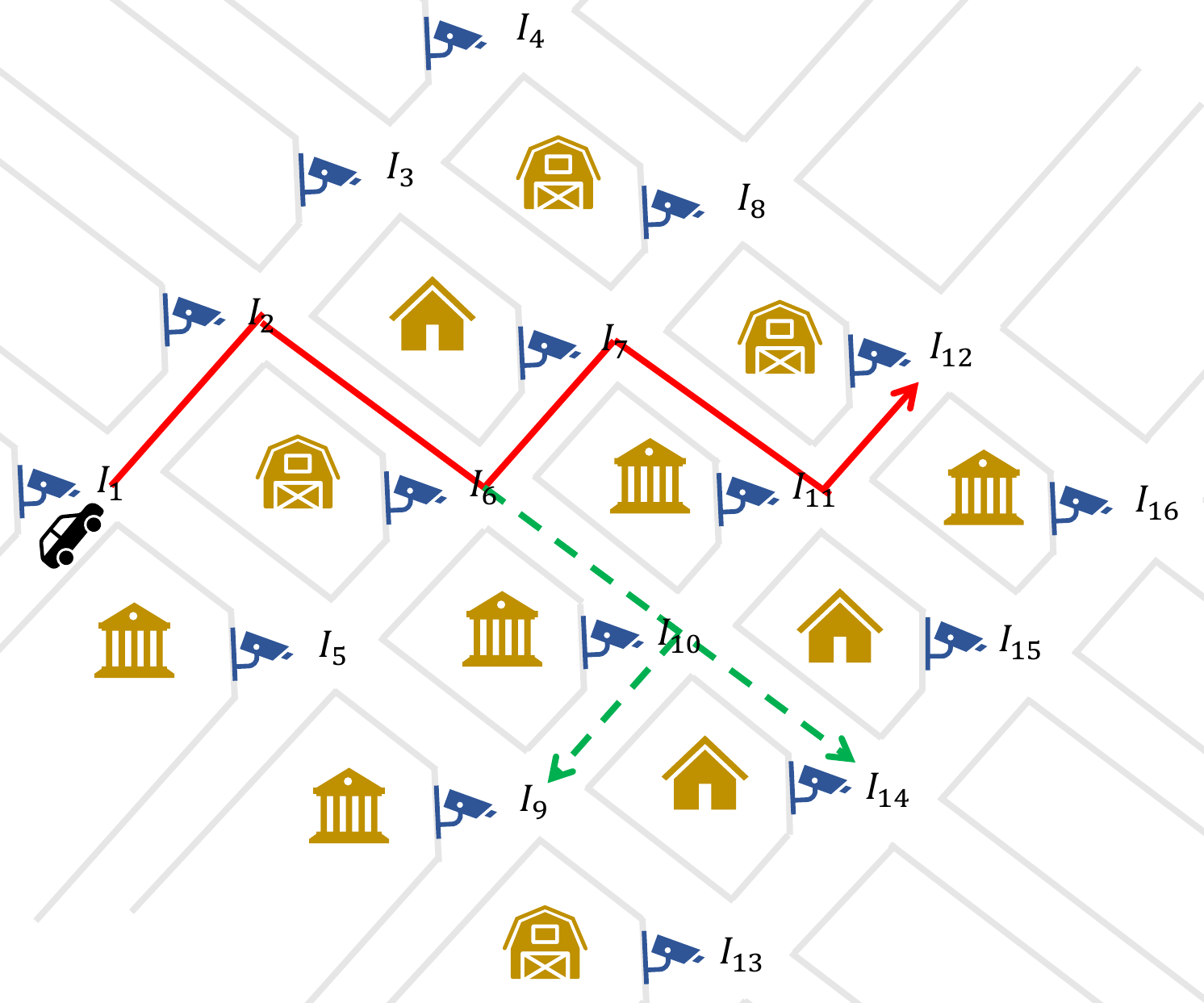}
\caption{\small An illustration example for tracking an VoI.}
\label{fig_watchdaog}
\vspace{-15pt}
\end{figure}

In light of WatchDog's architecture on each edge node, the tracking behavior of WatchDog can be described as follows:

\begin{enumerate}
    \item When a hit-and-run accident is reported at time instant $t^s_o$ from intersection $I_o$, WatchDog activates the first edge node deployed at $I_o$ to track the VoI and its active period is $[t_o^s, t_o^e]$. According to the number of vehicles detected at $I_o$, the proper machine learning modules are selected. By performing the real-time machine learning tasks on the detected vehicles, all the suspected vehicles and the next intersections on these vehicles' trajectories are identified. If $I_o$ is an uncrowded intersection, the VoI and the next intersection on the VoI's trajectory are identified. Then WatchDog actives all the edge nodes at the next intersections according to their active periods. Again, if $I_o$ is an uncrowded intersection, then only one next edge node will be activated.  
    \item The edge node deployed at $I_x$ is activated to track the VoI if the VoI or some suspected vehicles are identified at its neighboring intersection $I_{x-1}$ and traveling to $I_x$. The edge node's active period is $[t_x^s, t_x^e]$, which is calculated based on the active period of the previous edge node. The calculation method is introduced in Sec.~\ref{sec:active}. Again, similar to the first edge node, the proper machine learning modules are selected and performed to track all suspected vehicles traveling through this intersection during its active period.  
    \item Once the VoI is identified at any uncrowded intersection, all the suspected tracking branches are terminated.
\end{enumerate}

Step 2 and Step 3 are repeated iteratively to track the VoI in real-time.

\begin{example}
We use a simple example to illustrate how the whole tracking system works in Fig.~\ref{fig_watchdaog}. Suppose a Mercedes silver GLB SUV is reported to be involved in a hit-and-run accident at intersection $I_1$ at time instant $t^s_1$. The edge node at $I_1$ is activated to track the VoI. According to the number of vehicles detected from each frame, the admission control algorithm selects the most granular algorithm for the real-time ReID tasks to track the VoI and the VoI is found to travel to $I_2$. Then the edge node at $I_2$ is activated during to its active period. $I_2$ is an uncrowded intersection and the VoI is found to travel to $I_6$. However, $I_6$ is a crowded intersection and according to the number of vehicles detected from each frame, the admission control model selects color matching module for the real-time ReID tasks to track all the silver vehicles traveling through $I_6$. At $I_6$, some silver vehicles are found to travel to $I_7$ and the others are found to travel to $I_{10}$. Then, the edge nodes at $I_7$ and $I_{10}$ are activated to track all the silver vehicles. $I_{10}$ is anther crowded intersection and according to the number of vehicles detected from each frame, the admission control model selects color matching and model/make matching for the ReID tasks to track all suspected vehicles. Fortunately, $I_7$ is an uncrowded intersection and the VoI is identified at $I_7$, then all the other suspected tracking branches are terminated. The following involved edge nodes are activated in the same way to track the VoI in real-time. Note that a corner case is discussed in the captain of Fig.~\ref{fig_multiple}. 
\end{example}

\begin{figure}[t!]
\centering
\includegraphics[width=2.5in]{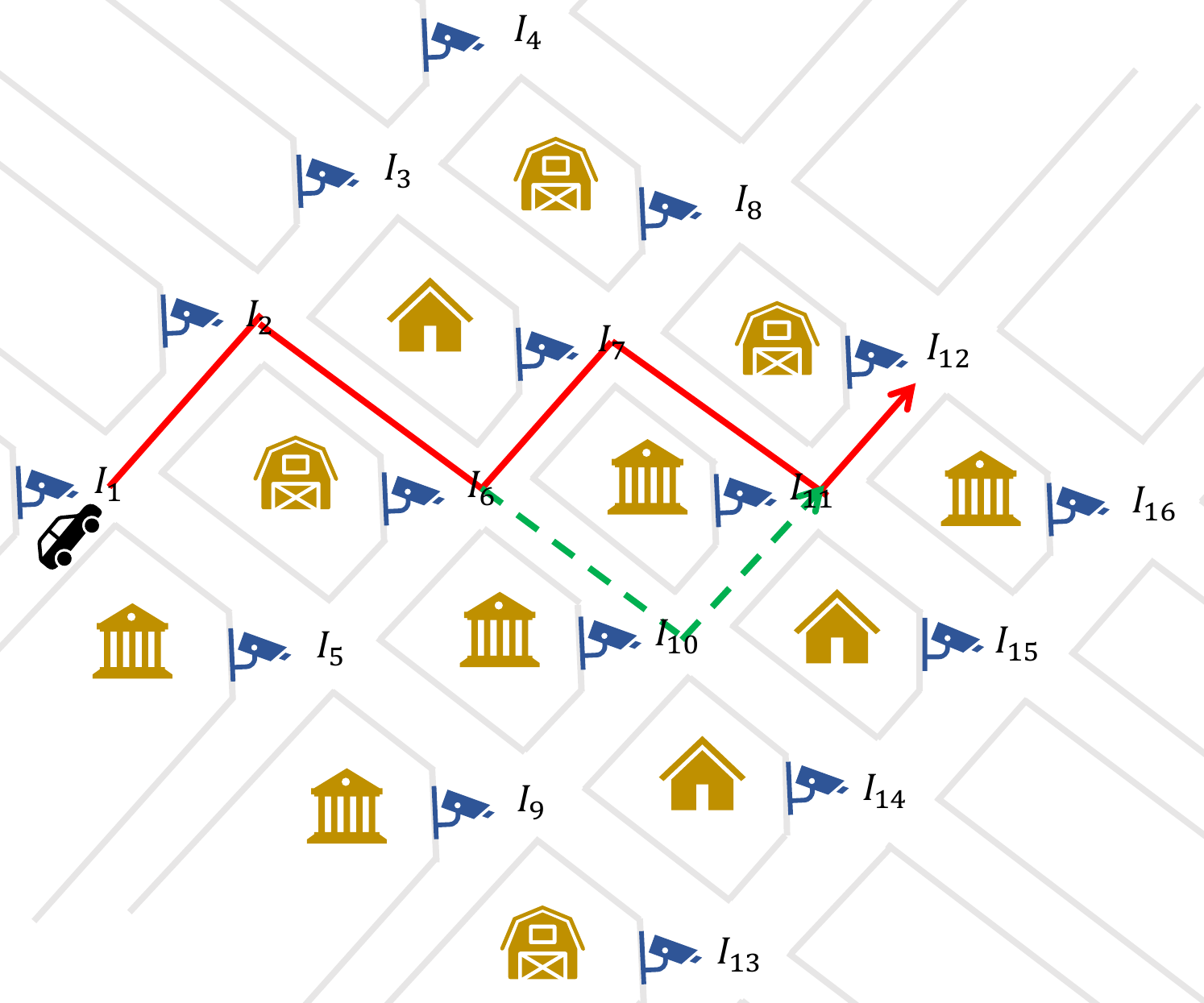}
\caption{\small The edge node at $I_{11}$ is activated to track the VoI by multiple edge nodes deployed at neighboring intersections $I_7$ and $I_{10}$. Since the active period of an edge node is calculated based on the active period of its previous edge node, the edge node at $I_{11}$ may have multiple active periods. If these periods overlap with each other, then the edge node's active period is a union of them.}
\label{fig_multiple}
\vspace{-15pt}
\end{figure}

Based on the above discussion, WatchDog can guarantee 100\% tracking coverage of the VoI without ``tracking loss''. We sacrifice the accuracy of ReID algorithms and track all the suspected vehicles at crowded intersections, then identify the VoI through utilizing the computing resource on the edge nodes deployed at uncrowded intersections. In a word, we develop a smart tracking method to make up for the limited computing capacity of a single edge node.

\section{Trace-Driven Evaluation}

To evaluate the efficacy of WatchDog in practice, we conducted extensive experiments based on our accessible real-world datasets.

\subsection{Data Description and Time/Traffic Measurement}\label{datasets}

Table~\ref{tab:my-table} summarizes statistics about vehicle networks studied in this work. To test WatchDog in a real-world scenario, we utilize a real-world dataset about 6 months of GPS traces of more than 14000 vehicles in Shenzhen, a Chinese city with 10 million population. The dataset is obtained by letting every vehicle upload its GPS records (the format as in Table~\ref{tab:my-table2}) to report its traces to a base station. Based on the dataset of GPS records, we obtain location and time distributions of the vehicles traveling in Shenzhen, which are used to evaluate the performance of WatchDog.

%\begin{figure*}
%\centering
%{
%\includegraphics[width=0.66\columnwidth]{Figs/trackingdelay.pdf}}\vspace{-5pt}
%{\label{fig:exp2}
%\includegraphics[width=0.66\columnwidth]{Figs/edgenodes.pdf}}\vspace{-5pt}
%{\label{fig:exp3}
%\includegraphics[width=0.7\columnwidth]{Figs/cost.pdf}}\vspace{-5pt}
%{\label{fig:exp4}}
%\caption{\footnotesize{The performance of Real-time tracking.}}\label{exp:all1}
%\end{figure*}

\begin{figure*}[!t]
	\centering
	\captionsetup[subfigure]{width = 0.9\columnwidth}
	\subfloat[Real-time Tracking Delay.]{\label{fig:exp2}
		\includegraphics[width=0.6\columnwidth]{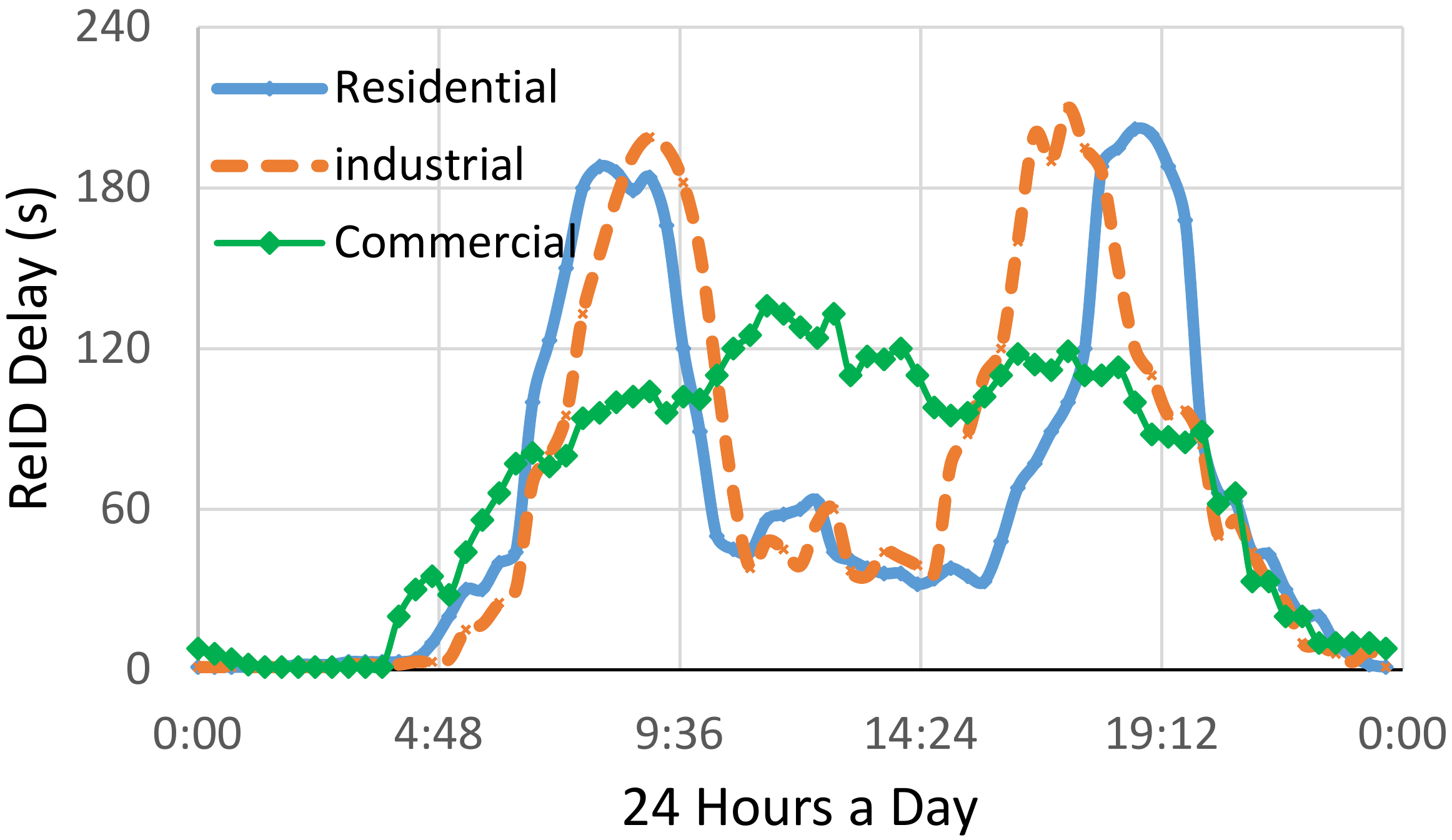}
	}\hfil
	\subfloat[The number of involved edge node to ReID a VoI.]{\label{fig:exp3}
		\includegraphics[width=0.6\columnwidth]{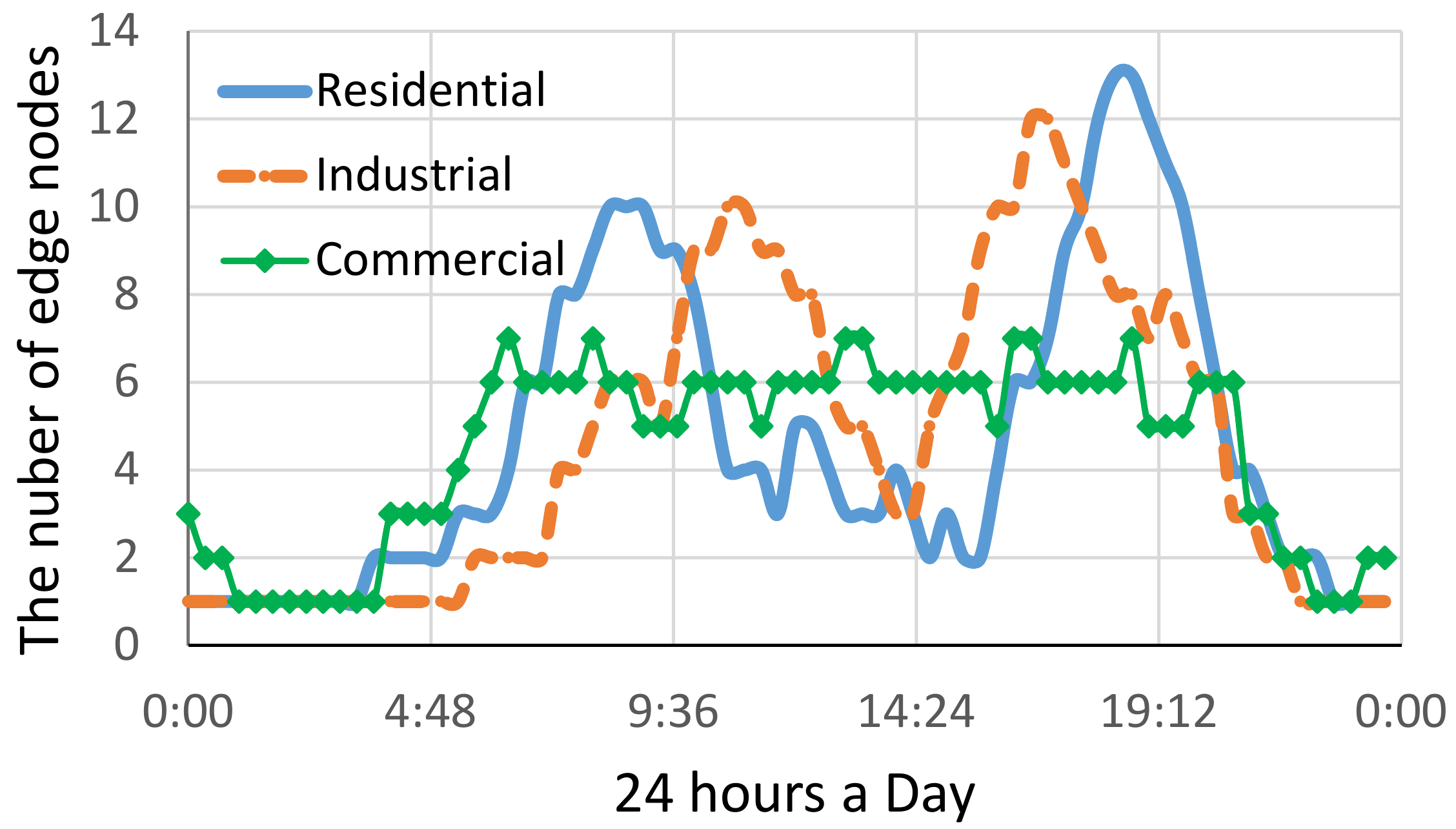}
	}\hfil
	\subfloat[The total execution time to ReID a VoI.]{\label{fig:exp4}
		\includegraphics[width=0.6\columnwidth]{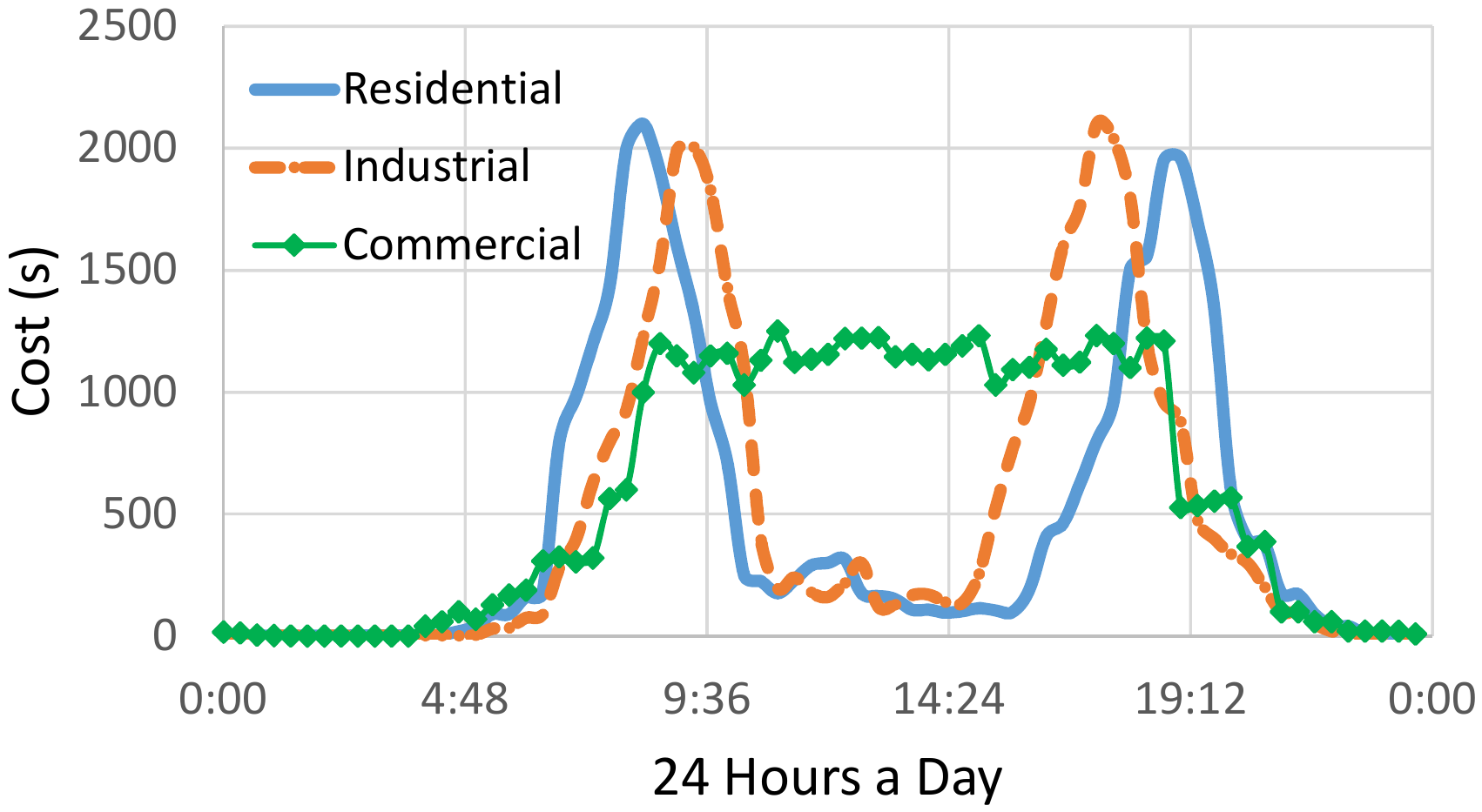}
	}
\caption{\small The performance of Real-time tracking.}\label{exp:all1}
\vspace{-10pt}
\end{figure*}

\begin{figure*}[!t]
	\centering
	\captionsetup[subfigure]{width = 0.9\columnwidth}
	\subfloat[Multi-Object Real-time Tracking Delay.]{\label{fig:exp5}
		\includegraphics[width=0.6\columnwidth]{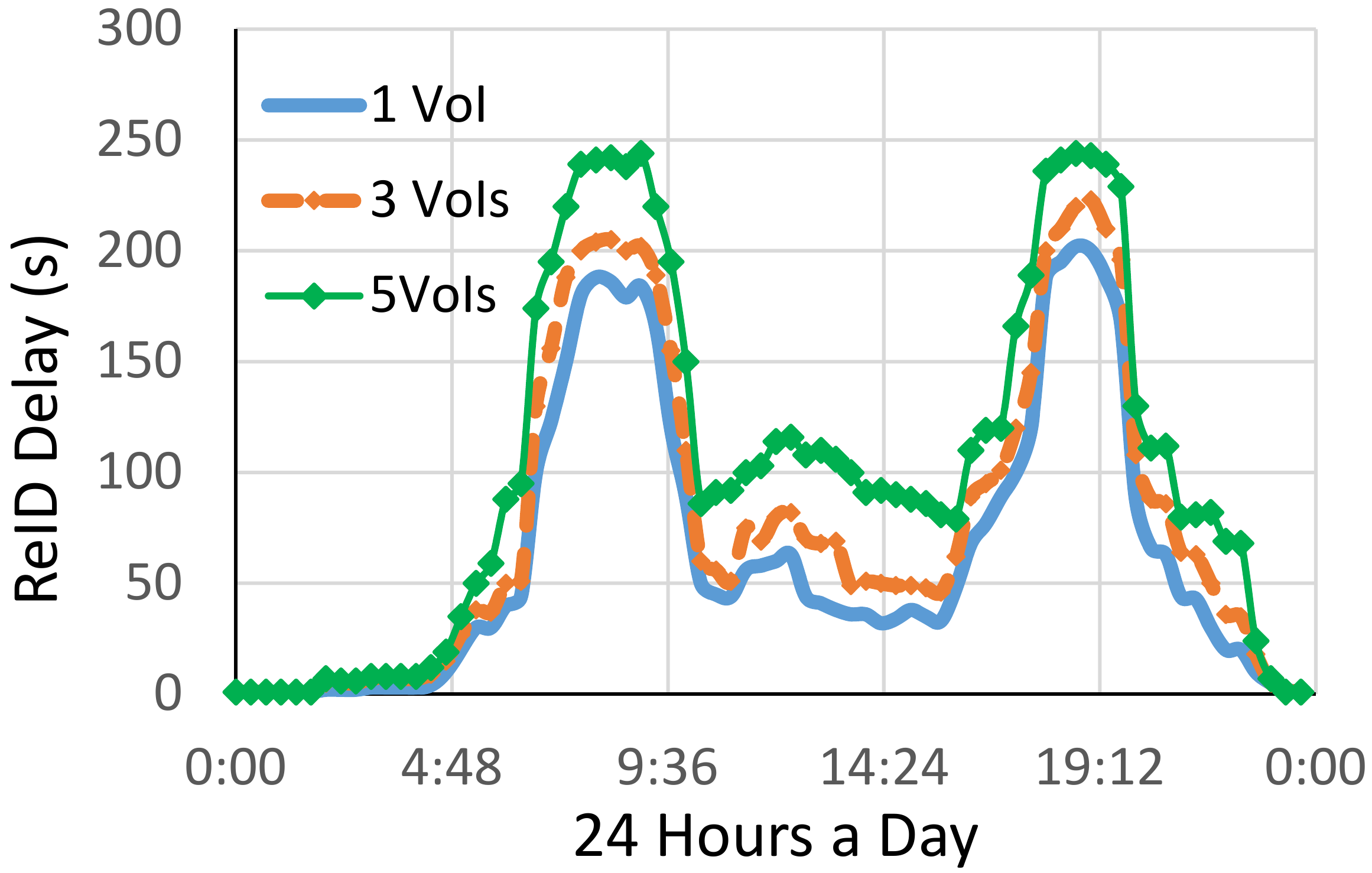}
	}\hfil
	\subfloat[The number of edge nodes to ReID multiple VoI.]{\label{fig:exp6}
		\includegraphics[width=0.6\columnwidth]{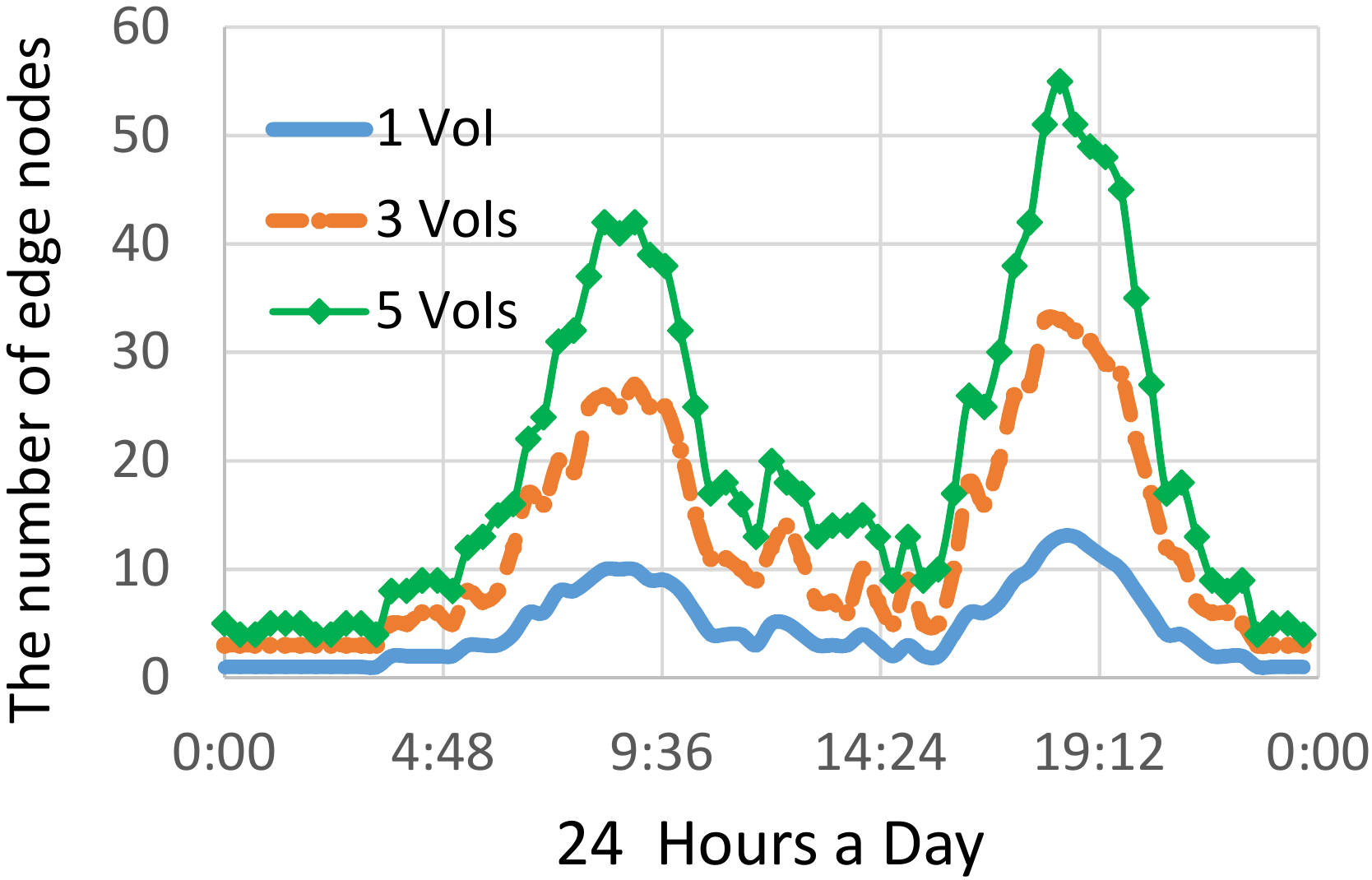}
	}\hfil
	\subfloat[The total execution time to ReID multiple VoIs.]{\label{fig:exp7}
		\includegraphics[width=0.6\columnwidth]{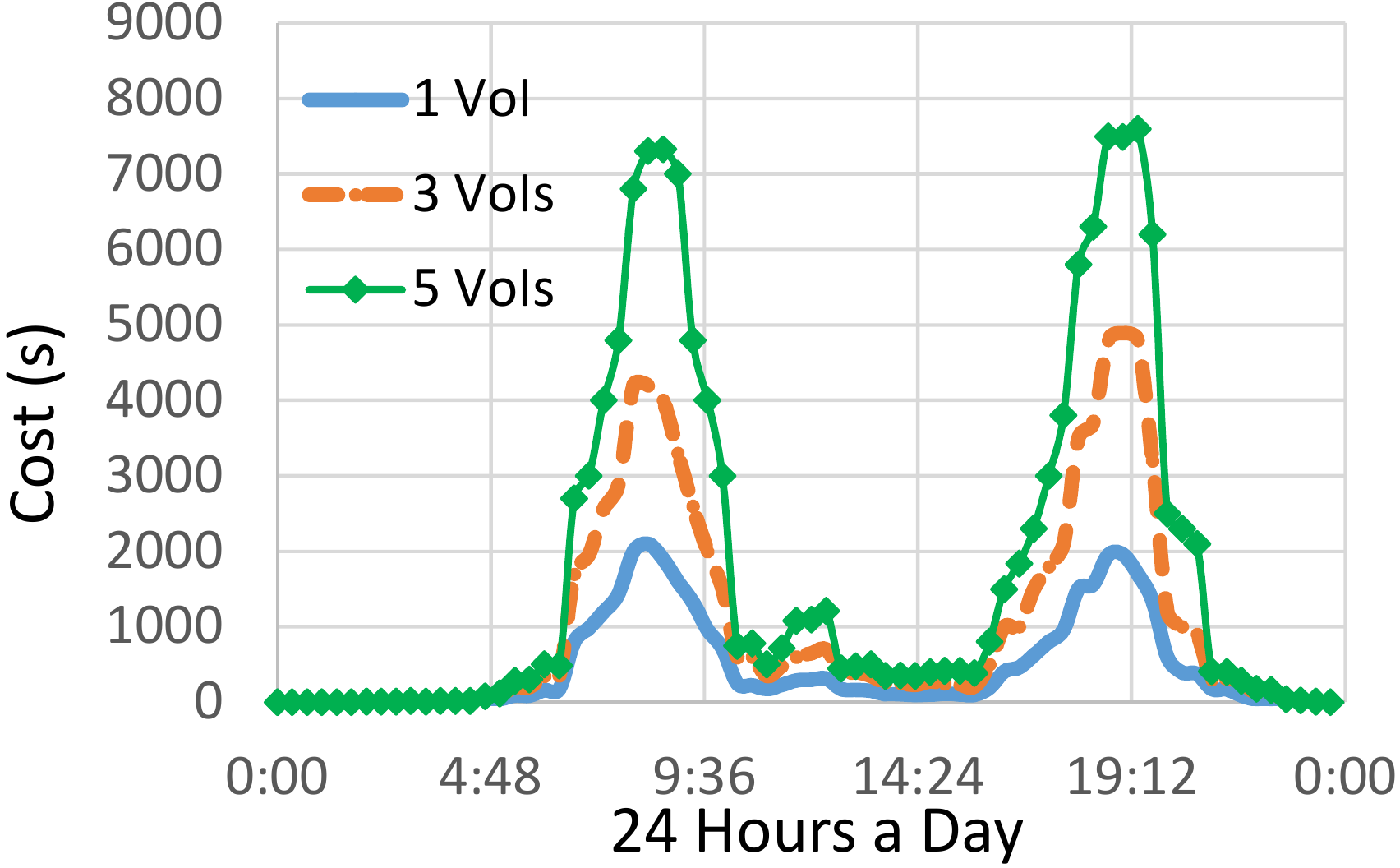}
	}
\caption{\small The performance of Multi-Object Real-time tracking.}\label{exp:all2}
\vspace{-10pt}
\end{figure*}

%\begin{figure*}
%\centering
%{
%\includegraphics[width=0.66\columnwidth]{Figs/multiDelay.pdf}}\vspace{-5pt}
%{\label{fig:exp2}
%\includegraphics[width=0.66\columnwidth]{Figs/multinodes.pdf}}\vspace{-5pt}
%{\label{fig:exp3}
%\includegraphics[width=0.7\columnwidth]{Figs/multiCost.pdf}}\vspace{-5pt}
%{\label{fig:exp4}}
%\caption{\footnotesize{The performance of Multi-Object Real-time %tracking.}}\label{exp:all2}
%\vspace{-10pt}
%\end{figure*}

\subsubsection{Measurement of traveling time}

The GPS data is used to measure the traveling time the VoI at specific locations in the monitored areas, including intersections and road segments. Fig.~\ref{fig:travellingtime}.(a) shows an example of traveling time taken by vehicles at an intersection. We can see that most of the vehicles take 5 seconds or 40 seconds to travel through this intersection. The reason is that if the traffic light at this intersection is red, the traveling time of a vehicle should include the waiting time. Due to different traveling speeds, the shortest time to travel through this intersection is 3 seconds and the longest time is 42 seconds according to Fig.~\ref{fig:travellingtime}.(a). Thus, we assume that the traveling time taken by the VoI at this intersection falls into this time interval, which is from 3 seconds to 42 seconds. Similarly, according to Fig.~\ref{fig:travellingtime}.(b), the traveling time taken by the VoI at the road segment belongs to a time interval, which is from 30 seconds to 50 seconds. By analyzing the GPS dataset, we can obtain the traveling time for the VoI at all intersections and road segments.

\begin{table}[]
\centering
\caption{Statistics of Vehicle Network}
\label{tab:my-table}
\begin{tabular}{|l|l|}
\hline
\multicolumn{2}{|c|}{Dataset Summary} \\ \hline
Collection Period         & 6 Months         \\ \hline
Collection Date         & 01/01/12-06/30/12          \\ \hline
Number of Vehicles         & 14,453         \\ \hline
Total Live Mile           & 371,269,642 miles      \\ \hline
\end{tabular}
\vspace{-15pt}
\end{table}

\begin{table}[h]
\caption{A GPS record.}
\label{tab:my-table2}
\begin{center}
\begin{tabular}{|c|c|c|c|c|}
\hline
%\textbf{PC}&\multicolumn{5}{|c|}{\textbf{Contribution Scores}} \\
%\cline{2-6} 
\textbf{plate ID} & \textbf{longitude}& \textbf{latitude}& \textbf{time}& \textbf{speed} \\
\hline
\hline
TIDXXXX & 114.022901 & 22.532104 & 08:34:43 & 22 km/h \\
\hline
\end{tabular}
\label{table}
\end{center}
\vspace{-10pt}
\end{table}

\subsubsection{Measurement of traffic condition}\label{sec:traffic} 

Based on the time and location information in each GPS record, the datasets can be used to measure the traffic conditions in the monitored area. Figure~\ref{fig_multiple} plots the percentage of the number of vehicles traveling through different intersections in one frame/one minute. At almost 16\% of the intersections, only nine vehicles appear in one minute. And for more than 95\% of the intersections, less than 21 vehicles travel through these intersections in one minute. These observations show that at most of the intersections, the traffic is light and our proposed method can fully utilize the edge nodes deployed at the uncrowded intersections to track the VoI in real-time. 

\begin{figure}[h]
\centering
\includegraphics[width=2.5in]{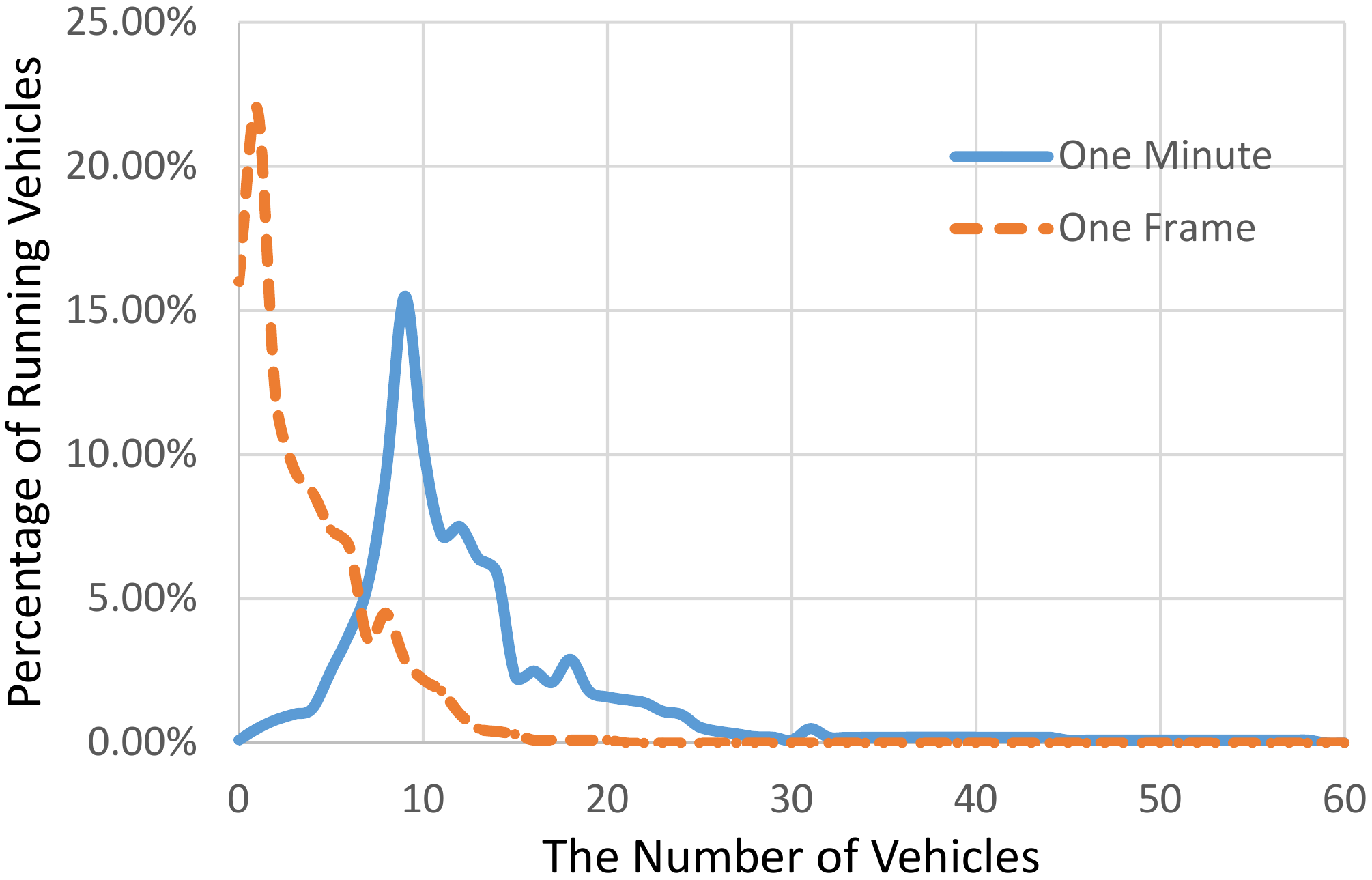}
\caption{\small The statistics of the number of vehicles traveling through an intersection in one minute/appearing on one video frame.}
\label{fig_multiple}
\vspace{-20pt}
\end{figure}

\subsection{Real-time Performance}\label{sec:performance}
The key performance metric for WatchDog is the video analytics delay to track the VoI in real-time. We evaluate this metric on every one hour time window of a day. In addition, we investigate the sensitivities
of WatchDog's performance on three key parameters, i.e., the tracking delay, the number of involved edged nodes, as well as the tracking cost.

In order to show the impacts of different traffic conditions on WatchDog, we evaluate the performance in three typical areas in Shenzhen: Residential area, Industrial area and Commercial area, which are denoted by ``Residential'', ``Industrial'' and ``Commercial'' in Fig.~\ref{exp:all1}, respectively. For each tracking experiment, we randomly choose a running vehicle from the dataset as the VoI and measure the tracking delay and tracking cost according to its driving circumstance at different intersections on its trajectory according to the GPS records. The ReID modules are selected by each involved edge node automatically, based on the real-time admission control module and the number of vehicles at the intersection. The execution time for each ReID module is given in Table~\ref{cost}. 

Fig.~\ref{fig:exp2} plots the average ReID delay during 24 hours of a day. During the rush hour, e.g., $06:00$-$10:00$, the ReID delay in three different areas reaches the maximal values. The reason is that the number of vehicles traveling through each intersection achieves the maximal value during the rush hour through the whole day. An interesting observation is that the peak of the Residential curve is earlier than the peak of the Industrial curve in the morning rush hour and on the contrary in the evening. This is because the traffic flows from residential area to industrial area in the morning and reversely in the evening. As seen in this figure, even during the rush hour, the ReID delay is smaller than 4 minutes, confirming that WatchDog can track the VoI in real-time.

Fig.~\ref{fig:exp3} plots the average number of edge nodes involved in tracking a VoI during 24 hours of a day. As seen in this figure, the maximum number of edge nodes involved in tracking a VoI is smaller than 14 even during the rush hour. This implies that through carefully selecting ReID modules for the edge nodes, WatchDog is able to efficiently reduce the number of suspected VoIs and limit the tracking range into a reasonably small area. More over, compared with residential area and industrial area, the number of involved edge nodes to track a VoI in commercial area in the rush hour is much smaller. This is because the workplaces for residents are distributed in the industrial area, thus the traffic in commercial area is not as heavy as that in residential/industrial areas in the rush hours.

We have also conducted a set of experiments evaluating the tracking cost in the three areas in terms of the total tracking time of all involved edge nodes to ReID a VoI during 24 hours of a day. This metric
can also be used to reflect the effectiveness of WatchDog since a shorter total tracking time to ReID and localize a VoI often results in a lower execution cost of the whole tracking system. Fig.~\ref{fig:exp4} shows the result using this metric. As seen, WatchDog can track the VoI with very few edge resources during non-rush hours and the cost for real-time tracking during the rush hour is also reasonable.

\subsection{Multi-Object Tracking}\label{sec:Multi-Object Tracking}

In this set of experiments, we evaluate the efficacy of WatchDog when handling the practical issue of multi-object real-time tracking (i.e., multiple hit-and-run accidents occur at the same time). Fig.~\ref{exp:all2} shows the evaluation results. We randomly choose multiple running vehicles (i.e., 3 and 5) as the VoIs and track them simultaneously according to the GPS datasets. We use the metric ``ReID delay'' (in seconds) to reflect the tracking latency of WatchDog. Similarly, we use the metrics, ``the number of invlved edge nodes'' and ``Cost'', to show the total execution time of the edge nodes consumed by WatchDog when tracking the multiple VoIs in real-time.

Fig.~\ref{exp:all2} shows the results w.r.t. the multi-object real-time tracking. As seen in the figure, with increased number of VoIs, all the measured parameters increase during all time intervals. This observation confirms the intuition that as more VoIs involved in the real-time tracking, the larger amount of edge resources is needed to track all the VoIs simultaneously. One interesting observation is that the amount of increased edge resources is not proportional to the number of increased VoIs. For example, in Fig.~\ref{fig:exp7}, the cost for tracking 1 VoI achieves its maximum value at around 8:00 AM, which is about 2000 seconds. However, the cost for tracking 5 VoIs at 8:00 AM is less than 8000 seconds. This implies that the intersections involved in tracking the 5 VoIs have overlaps with each other and the video frame processing results are reused to track different VoIs. Another observation in Fig.~\ref{fig:exp5} is that the ReID Delays for tracking different number of VoIs are very close. The reason is that our proposed WatchDog is a distributed real-time tracking system, which can perform the multi-object tacking simultaneously.    
\section{Conclusion}
Recent technology advances in edge computing provide new opportunities to implement a real-time tracking system in smart cities with edge nodes distributed at the intersections of the road network, which consist of both surveillance cameras and embedded computing platforms. We propose a simple yet effective real-time system for tracking hit-and-run vehicles in smart cities, which employs machine learning tasks with different resource-accuracy trade-offs, and schedule tracking tasks across distributed edge nodes based on the number of detected vehicles to maximize the execution time of tasks while ensuring a provable completion time bound at each edge node. WatchDog is also designed to be
capable of addressing multi-object tracking problem to track multiple VoIs simultaneously in real-time.

\bibliographystyle{IEEEtran}
\bibliography{WatchDog}

\begin{thebibliography}{10}
\providecommand{\url}[1]{#1}
\csname url@samestyle\endcsname
\providecommand{\newblock}{\relax}
\providecommand{\bibinfo}[2]{#2}
\providecommand{\BIBentrySTDinterwordspacing}{\spaceskip=0pt\relax}
\providecommand{\BIBentryALTinterwordstretchfactor}{4}
\providecommand{\BIBentryALTinterwordspacing}{\spaceskip=\fontdimen2\font plus
\BIBentryALTinterwordstretchfactor\fontdimen3\font minus
  \fontdimen4\font\relax}
\providecommand{\BIBforeignlanguage}[2]{{%
\expandafter\ifx\csname l@#1\endcsname\relax
\typeout{** WARNING: IEEEtran.bst: No hyphenation pattern has been}%
\typeout{** loaded for the language `#1'. Using the pattern for}%
\typeout{** the default language instead.}%
\else
\language=\csname l@#1\endcsname
\fi
#2}}
\providecommand{\BIBdecl}{\relax}
\BIBdecl

\bibitem{Azure}
\emph{Microsoft Azure Stack Edge}, \url{https://azure.microsoft.com}.

\bibitem{leduc2008road}
G.~Leduc \emph{et~al.}, ``Road traffic data: Collection methods and
  applications,'' \emph{Working Papers on Energy, Transport and Climate Change
  2008}.

\bibitem{bramberger2004real}
M.~Bramberger, J.~Brunner, B.~Rinner, and H.~Schwabach, ``Real-time video
  analysis on an embedded smart camera for traffic surveillance,'' in
  \emph{RTAS 2004}.

\bibitem{ishii2005monitor}
Y.~Ishii, ``Monitor system for monitoring suspicious object,'' Dec.~15 2005, uS
  Patent App. 11/150,264.

\bibitem{hung2018videoedge}
C.-C. Hung, G.~Ananthanarayanan, P.~Bodík, L.~Golubchik, M.~Yu, V.~Bahl, and
  M.~Philipose, ``{VideoEdge: Processing Camera Streams using Hierarchical
  Clusters},'' in \emph{ACM/IEEE SEC}, 2018.

\bibitem{wang19hotedgevideo}
J.~Jiang, Y.~Zhou, G.~Ananthanarayanan, Y.~Shu, and A.~A. Chien, in \emph{Hot
  edge 2019}.

\bibitem{ross2015RCNN}
R.~Girshick, ``Fast r-cnn,'' in \emph{ICCV 2015}.

\bibitem{ren2015FRCNN}
R.~Shaoqing, H.~Kaiming, G.~Ross, and S.~Jian, ``Faster r-cnn: Towards
  real-time object detection with region proposal networks,'' in \emph{NIPS
  2015}.

\bibitem{dai2016RFCNN}
D.~Jifeng, L.~Yi, H.~Kaiming, and S.~Jian, ``R-fcn: Object detection via
  region-based fully convolutional networks,'' in \emph{NIPS 2016}.

\bibitem{he2017maskrcnn}
H.~Kaiming, Georgia, and D.~Piotr, ``Mask r-cnn,'' in \emph{ICCV 2017}.

\bibitem{boxy2019}
K.~Behrendt, ``Boxy vehicle detection in large images,'' in \emph{ICCV 2019}.

\bibitem{marius2016cityscapes}
C.~Marius, O.~Mohamed, R.~Sebastian, R.~Timo, E.~Markus, B.~Rodrigo, F.~Uwe,
  R.~Stefan, and S.~Bernt, ``Feature pyramid networks for object detection,''
  in \emph{CVPR 2016}.

\bibitem{geiger2012CVPR}
G.~Andreas, L.~Philip, and U.~Raquel, ``Are we ready for autonomous driving?
  the kitti vision benchmark suite,'' in \emph{CVPR}.\hskip 1em plus 0.5em
  minus 0.4em\relax IEEE, 2012.

\bibitem{redmon2016yolov1}
R.~Joseph, D.~Santosh, G.~Ross, and F.~Ali, ``You only look once: Unified,
  real-time object detection,'' in \emph{CVPR 2016}.

\bibitem{liu2016ssd}
L.~Wei, A.~Dragomir, E.~Dumitru, S.~Christian, R.~Scott, F.~Cheng-Yang, and
  C.~B. Alexander, ``Ssd: Single shot multibox detector,'' in \emph{European
  Conference on Computer Vision (ECCV)}.\hskip 1em plus 0.5em minus 0.4em\relax
  Springer, 2016.

\bibitem{lin2017RetinaNet}
L.~Tsung-Yi, G.~Priya, G.~Ross, H.~Kaiming, and D.~Piotr, ``Focal loss for
  dense object detection,'' in \emph{ICCV 2017}.

\bibitem{li2018yolov3}
R.~Joseph and F.~Ali, ``Yolov3: An incremental improvement,'' \emph{arXiv
  preprint arXiv:1804.02767}, 2018.

\bibitem{khorramshahi2019vreid}
K.~Pirazh, P.~Neehar, K.~Amit, S.~Anshul, and C.~Rama, ``Attention driven
  vehicle re-identification and unsupervised anomaly detection for traffic
  understanding,'' in \emph{CVPR 2019}.

\bibitem{tan2019vreid}
T.~Xiao, W.~Zhigang, J.~Minyue, Y.~Xipeng, W.~Jian, G.~Yuan, S.~Xiangbo,
  Y.~Xiaoqing, Y.~Yuchen, H.~Dongliang, W.~Shilei, and D.~Errui, ``Multi-camera
  vehicle tracking and re-identification based on visual and spatial-temporal
  features,'' in \emph{CVPR 2019}.

\bibitem{huang2019vreid}
H.~Tsung-Wei, C.~Jiarui, Y.~Hao, H.~Hung-Min, and H.~Jenq-Neng, ``Multi-view
  vehicle re-identification using temporal attention model and metadata
  re-ranking,'' in \emph{CVPR 2019}.

\bibitem{spanhel2019vreid}
S.~Jakub, B.~Vojtech, and H.~Adam, ``Vehicle re-identifiation and multi-camera
  tracking in challenging city-scale environment,'' in \emph{CVPR 2019}.

\bibitem{hsu2019vreid}
H.~Hung-Min, H.~Tsung-Wei, W.~Gaoang, C.~Jiarui, L.~Zhichao, and H.~Jenq-Neng,
  ``Multi-camera tracking of vehicles based on deep features re-id and
  trajectory-based camera link models,'' in \emph{CVPR 2019}.

\bibitem{gonzales2002digital}
R.~C. Gonzales and R.~E. Woods, ``Digital image processing,'' 2002.

\bibitem{guru2010knn}
G.~DS, Y.~Sharath, and S.~Manjunath, ``Texture features and knn in
  classification of flower images,'' \emph{IJCA 2010}.

\bibitem{david1999SIFT}
D.~G. Lowe, ``Object recognition from local scale-invariant features,'' in
  \emph{ICCV 1999}.

\bibitem{herbert2008surf}
B.~Herbert, E.~Andreas, T.~Tinne, and V.~G. Luc, ``Speeded-up robust features
  (surf),'' \emph{CVIU 2008}.

\bibitem{dalal2005hog}
D.~Navneet and T.~Bill, ``Histograms of oriented gradients for human
  detection,'' in \emph{CVPR 2005}.

\bibitem{baran2015SVM}
B.~Remigiusz, G.~Andrzej, and M.~Andrzej, ``The efficient real- and
  non-real-time make and model recognition of cars,'' \emph{MTA 2015}.

\bibitem{noppakun2017svm}
B.~Noppakun and P.~Simant, ``Car make and model recognition under limited
  lighting conditions at night,'' \emph{CVIU 2017}.

\bibitem{galiano2012random}
V.F.Rodriguez-Galiano, B.Ghimire, J.Rogan, M.Chica-Olmo, and J.P.Rigol-Sanchez,
  ``An assessment of the effectiveness of a random forest classifier for
  land-cover classification,'' \emph{ISPRS 2012}.

\bibitem{faezeh2017deep}
T.~Faezeh, F.~Hichem, and N.~Keishin, ``A large and diverse dataset for
  improved vehicle make and model recognition,'' in \emph{CVPR 2017}.

\bibitem{jo2018deep}
S.~Y. Jo, N.~Ahn, Y.~Lee, and S.-J. Kang, ``Transfer learning-based vehicle
  classification,'' in \emph{ISOCC 2018}.

\bibitem{he2016deep}
K.~He, X.~Zhang, S.~Ren, and J.~Sun, ``Deep residual learning for image
  recognition,'' in \emph{CVPR 2016}.

\bibitem{sandler2018mobilenetv2}
M.~Sandler, A.~Howard, M.~Zhu, A.~Zhmoginov, and L.-C. Chen, ``Mobilenetv2:
  Inverted residuals and linear bottlenecks,'' in \emph{CVPR 2018}.

\bibitem{zhang2018shufflenet}
X.~Zhang, X.~Zhou, and J.~Sun, ``Shufflenet: An extremely efficient
  convolutional neural network for mobile devices,'' in \emph{CVPR 2018}.

\bibitem{tan2019efficientnet}
M.~Tan and Q.~V. Le, ``Efficientnet: Rethinking model scaling for convolutional
  neural networks,'' in \emph{ICML 2019}.

\bibitem{david1999ensemble}
O.~David and M.~Richard, ``Popular ensemble methods: An empirical study,''
  \emph{JAIR 1999}.

\bibitem{pirazh2019vreid}
K.~Pirazh, K.~Amit, P.~Neehar, S.~S. Rambhatla, C.~Jun-Cheng, and C.~Rama, ``A
  dual-path model with adaptive attention for vehicle re-identification,'' in
  \emph{ICCV 2019}.

\bibitem{he2019vreid}
H.~Bing, L.~Jia, Z.~Yifan, and T.~Yonghong, ``Part-regularized near-duplicate
  vehicle re-identification,'' in \emph{CVPR 2019}.

\bibitem{wang2017vreid}
W.~Zhongdao, T.~Luming, L.~Xihui, Y.~Zhuliang, Y.~Shuai, S.~Jing, Y.~Junjie,
  W.~Shengjin, L.~Hongsheng, and W.~Xiaogang, ``Orientation invariant feature
  embedding and spatial temporal regularization for vehicle reidentification,''
  in \emph{ICCV 2017}.

\bibitem{georgia2019vreid}
R.~Georgia, K.~Aytac, L.~Minxian, and G.~Shaogang, ``Multi-task mutual learning
  for vehicle re-identification,'' in \emph{CVPR 2019}.

\bibitem{zhang2018multitask}
Z.~Yu, W.~Ying, and Y.~Qiang, ``Learning to multitask,'' in \emph{NeurIPS},
  2018.

\bibitem{chen2017multitask}
C.~Weihua, C.~Xiaotang, Z.~Jianguo, and H.~Kaiqi, ``A multi-task deep network
  for person re-identification,'' in \emph{AAAI 2017}.

\bibitem{wang2018granularities}
W.~Guanshuo, Y.~Yufeng, C.~Xiong, L.~Jiwei, and Z.~Xi, ``Learning
  discriminative features with multiple granularities for person
  re-identification,'' in \emph{MM 2018}.

\bibitem{hu2018senet}
H.~Jie, S.~Li, and S.~Gang, ``Squeeze-and-excitation networks.''\hskip 1em plus
  0.5em minus 0.4em\relax IEEE, 2018.

\bibitem{leontyev2007tardiness}
H.~Leontyev and J.~H. Anderson, ``Tardiness bounds for fifo scheduling on
  multiprocessors,'' in \emph{ECRTS'07}.

\bibitem{dong2018general}
Z.~Dong, C.~Liu, S.~Bateni, Z.~Kong, L.~He, L.~Zhang, R.~Prakash, and Y.~Zhang,
  ``A general analysis framework for soft real-time tasks,'' \emph{IEEE
  Transactions on Parallel and Distributed Systems}, vol.~30, no.~6, pp.
  1222--1237, 2018.

\end{thebibliography}

\end{document}